\documentclass[12pt]{article}%
\usepackage{amsmath,amssymb,amsthm,amsfonts}
\usepackage{wasysym}
\usepackage{subcaption}
\usepackage{graphicx}
\usepackage[colorlinks]{hyperref}
\usepackage{color}
\usepackage{geometry}
\usepackage{appendix}
\usepackage{textcomp}

\newtheorem{thm}{Theorem}
\theoremstyle{definition}

\newcommand{\R}{\mathbb{R}}

\newcommand{\N}{\mathbb{N}}
\newcommand{\card}{\mathbf{card}}

\newcommand{\ea}{\textit{et al. }}
\newcommand{\nor}{\,\bar{\vee}\,}

\renewcommand{\setminus}{\smallsetminus}

\begin{document}

\title{Qualitative models and experimental investigation of chaotic NOR gates and set/reset flip-flops}

\author{Aminur Rahman\thanks{Corresponding Author, \url{ar276@njit.edu}}
\thanks{Department of Mathematical Sciences, New Jersey Institute of Technology} ,
Ian Jordan\thanks{Department of Electrical and Computer Engineering, New Jersey Institute of Technology} ,
Denis Blackmore\footnotemark[2]}

\date{}
\maketitle

\begin{abstract}
It has been observed through experiments and \emph{SPICE} simulations that logical circuits based
upon Chua's circuit exhibit complex dynamical behavior.  This behavior can be used to design analogs of more
complex logic families and some properties can be exploited for electronics applications.  Some of these circuits have been
modeled as systems of ordinary differential equations.  However, as the number of components
in newer circuits increases so does the complexity.  This renders continuous dynamical
systems models impractical and necessitates new modeling techniques.  In recent
years some discrete dynamical models have been developed using various
simplifying assumptions.  To create a robust modeling framework for chaotic logical circuits,
we developed both deterministic and stochastic discrete dynamical models, which exploit
the natural recurrence behavior, for two chaotic NOR gates and a chaotic set/reset flip-flop (RSFF).
This work presents a complete applied mathematical investigation of logical circuits.  Experiments
on our own designs of the above circuits are modeled and the models are rigorously analyzed
and simulated showing surprisingly close qualitative agreement with the experiments.  Furthermore,
the models are designed to accommodate dynamics of similarly designed circuits.  This
will allow researchers to develop ever more complex chaotic logical circuits with a simple
modeling framework.
\end{abstract}


Pacs numbers: 05.45.Ac, 
05.45.Pq, 
84.30.-r 

\maketitle

\section{Introduction}

Since the 1990s there has been growing interest in controlling chaotic circuits
starting from synchronization \cite{Pecora-Carrol90, Cuomo-Oppenheim93},
and leading to logical circuits \cite{ONK1998}.
Constructions of chaotic logical circuits mainly employ the usual circuit elements
such as resistors, capacitors, and inductors, and a less common component called
a \emph{nonlinear resistor}.  The most well-known nonlinear resistor is Chua's diode,
invented by Leon Chua in 1983 and used as an integral element in Chua's circuit
\cite{Matsumoto84, Kennedy92, CWHZ93, Chua94}.  For more complex logical
circuits, components called \emph{threshold control units} (TCUs)\cite{Murali-Sinha03} have been employed
\cite{MSD2003A, MSD2003B, Cafagna-Grassi06}.

This may seem counterintuitive since throughout the latter half of the 20\textsuperscript{th}
century we have constructed ever more stable electronic components to be
used in computers and other devices.  This has been a triumph for electrical
engineering and physics.  However, not all logical systems are electronic nor
man-made \cite{HSWK14}, and as we may observe, nature is often unstable.  Since it is easier to
study electrical systems, these chaotic logical circuits can help us better
understand naturally occurring logical systems.  Furthermore, the chaotic/logical
properties of the circuits can be exploited for the purposes of encryption or
secure communication.  

While there has been an abundance of \emph{SPICE} simulations and some experimental investigations in the
literature, such as \cite{ONK1998, MSD2003A, MSD2003B}, there is a dearth
of models for the more complex chaotic logical
circuits.  This is understandable since the usual modeling techniques become
ever more difficult to implement as the number of components increase.
Traditionally, logical circuits have been modeled as systems of ordinary
differential equations (see \cite{Kacprzak88, ONK1998}) because resistors,
capacitors, and inductors are related via different rates of change.  Furthermore,
\emph{SPICE} simulations are often employed as a means of studying circuit designs,
but the computational costs become unreasonable as circuits become more complex.
However, more recently, there has been some effort in modeling logical circuits as simple
discrete dynamical systems \cite{HDJ92, RZW2006, BRS2009} and developing a
mathematical framework for studying chaos in boolean networks \cite{CGSZ10, RahmanBlackmore16}.

To facilitate the development of models of more complex chaotic logical circuits,
we created a modeling framework by investigating two chaotic NOR gates and a
chaotic Set/Reset flip-flop (RSFF).  We modified the chaotic RSFF/dual NOR gate
design of Cafagna and Grassi \cite{Cafagna-Grassi06} and simulated our design
in \emph{MultiSIM} (a software based on \emph{SPICE}) to verify agreement with the \emph{PSPICE}
simulations of \cite{Cafagna-Grassi06}.  Once the design was satisfactory, we built the
circuits and recorded the same measurements as the simulations for the sake of having
compatible data sets.  These empirical observations and information about the physics
from previous investigations were then used to develop the models.  This was followed
by analysis and simulations of our models, which showed surprising agreement with the
experiments and \emph{SPICE} based simulations.

\begin{figure}[htbp]
\centering
\includegraphics[width = 0.45\textwidth]{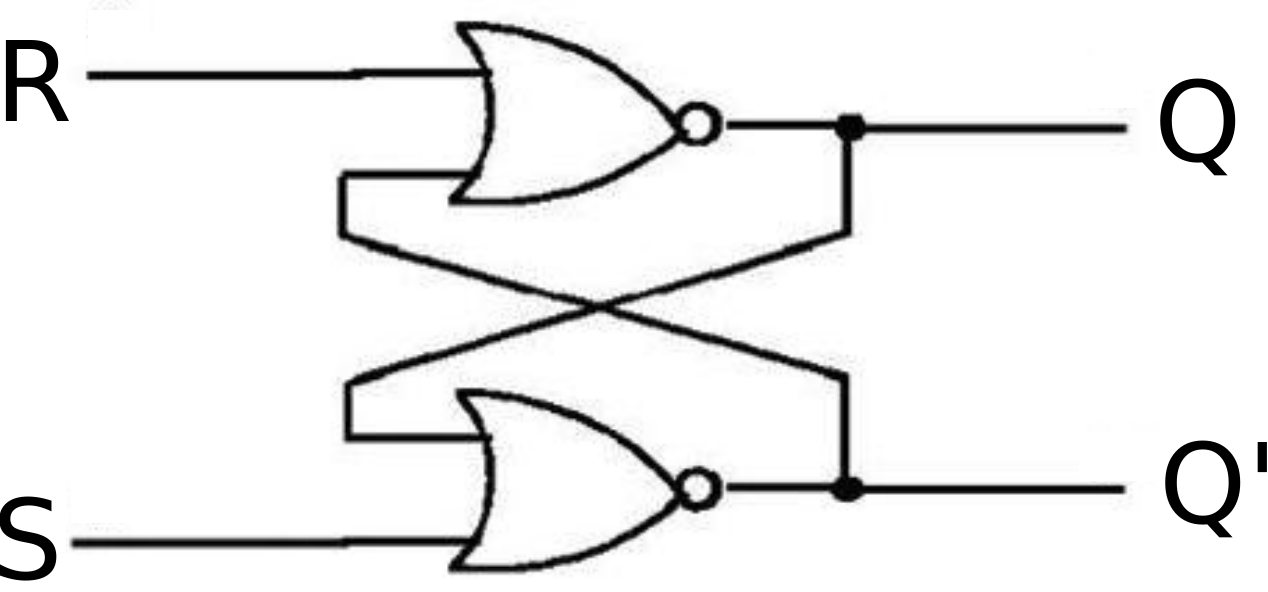}
\qquad
\large
\begin{tabular}[b]{|l|l|l|l|}
R & S & Q & Q'\\
\hline
0 & 0 & \multicolumn{2}{|c|}{Keep previous state}\\
\hline
0 & 1 & 1 & 0\\
\hline
1 & 0 & 0 & 1\\
\hline
1 & 1 & \multicolumn{2}{|c|}{Not allowed}\\
\hline
\end{tabular}
\caption{Black box schematic and table of operations for the Set/Reset flip-flop circuit.}\label{Fig: BlackBox}
\end{figure}

Since the full schematic of the circuits (3 in 1) is quite complex and similar to those in
the aforementioned works,
it have been relegated to the appendix.  However, the ``black box'' schematics of both
types of circuits are quite simple, and will be referred to throughout this work.
For example, a NOR gate acts as the negation of the disjunctive operator, i.e. it will output
high voltage if and only if it receives low voltage inputs.  The RSFF employs
two NOR gates with feedbacks as shown in Fig \ref{Fig: BlackBox}.

The body of this work is organized into two main parts: experiments and extensive modeling,
analysis, and simulations.
In section \ref{Sec: Experiments} we briefly refer to past \emph{SPICE} simulations and discuss
experimental results of our circuit designs in order to motivate the models.  Section \ref{Sec: Deterministic}
contains the focus of this article.  We first derive deterministic discrete
dynamical models for two types of chaotic NOR gates and analyze certain
properties of the models including the existence of chaotic dynamics.
Then we use the NOR gate models to derive the RSFF model and
compare their simulations with the experiments.  Next, in Section \ref{Sec: Stochastic}
we derive a stochastic discrete dynamical model for the RSFF in order to
incorporate \emph{races} (when two parallel signals do not have the same
``speed'') and compare their simulations with experiments.  We end by discussing
unexpected predictive capabilities of the models for components that were
not explicitly accounted for.

\section{Experiments and \emph{SPICE} Simulations}\label{Sec: Experiments}

In this section we discuss previous \emph{PSPICE} simulations for chaotic NOR gate and RSFF constructions
and present our own design and experimental results to demonstrate the robustness of the
circuit design even in the midst of chaos.

\subsection{Previous investigations}

Murali \ea developed and tested a chaotic NOR gate construction in \cite{MSD2003A}
and with more detail in \cite{MSD2003B}.  Then Cafagna and Grassi designed
an RSFF \cite{Cafagna-Grassi06} using similar principles to the NOR gate of Murali \ea,
which is now the inspiration for our own RSFF.

The RSFF is designed with Chua's circuit at its core and two TCUs
to realize each NOR gate.  Two switches are used to convert the circuit from operating
as two separate NOR gates to operating as an RSFF and vice versa.  Cafagna and Grassi
provide plots of the inputs, outputs, and threshold voltages for both NOR gates, the
inputs and outputs of the RSFF, and finally the voltages across the two capacitors.
We recreate the same types of plots for our design in order to facilitate comparisons.

\subsection{Set/reset flip-flop design and experiment}

\begin{figure}[htbp]
\begin{subfigure}{0.46\textwidth}
\includegraphics[width = \textwidth]{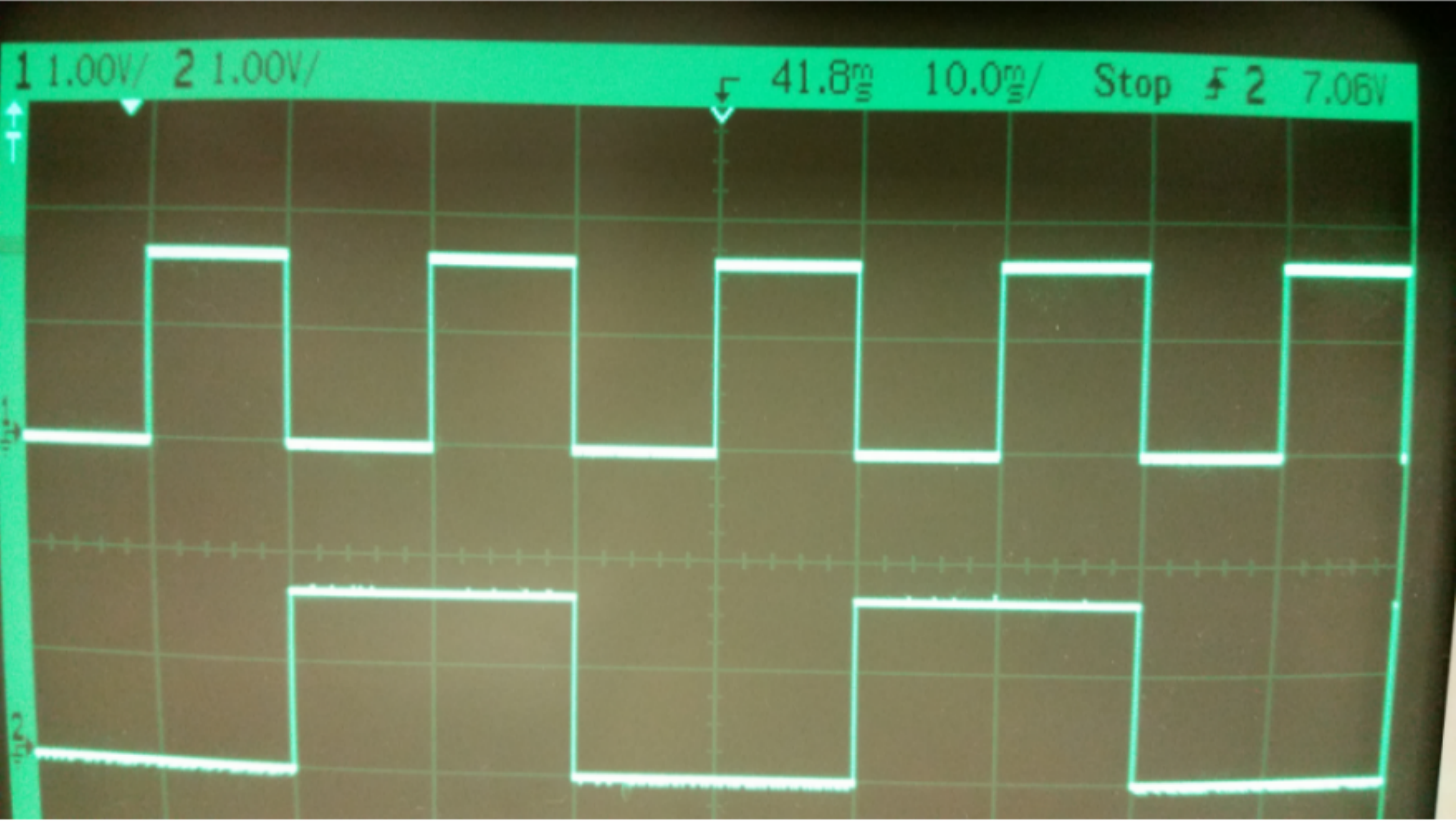}
\caption{Inputs}\label{Fig: ExpInputs}
\end{subfigure}
\begin{subfigure}{0.52\textwidth}
\includegraphics[width = \textwidth]{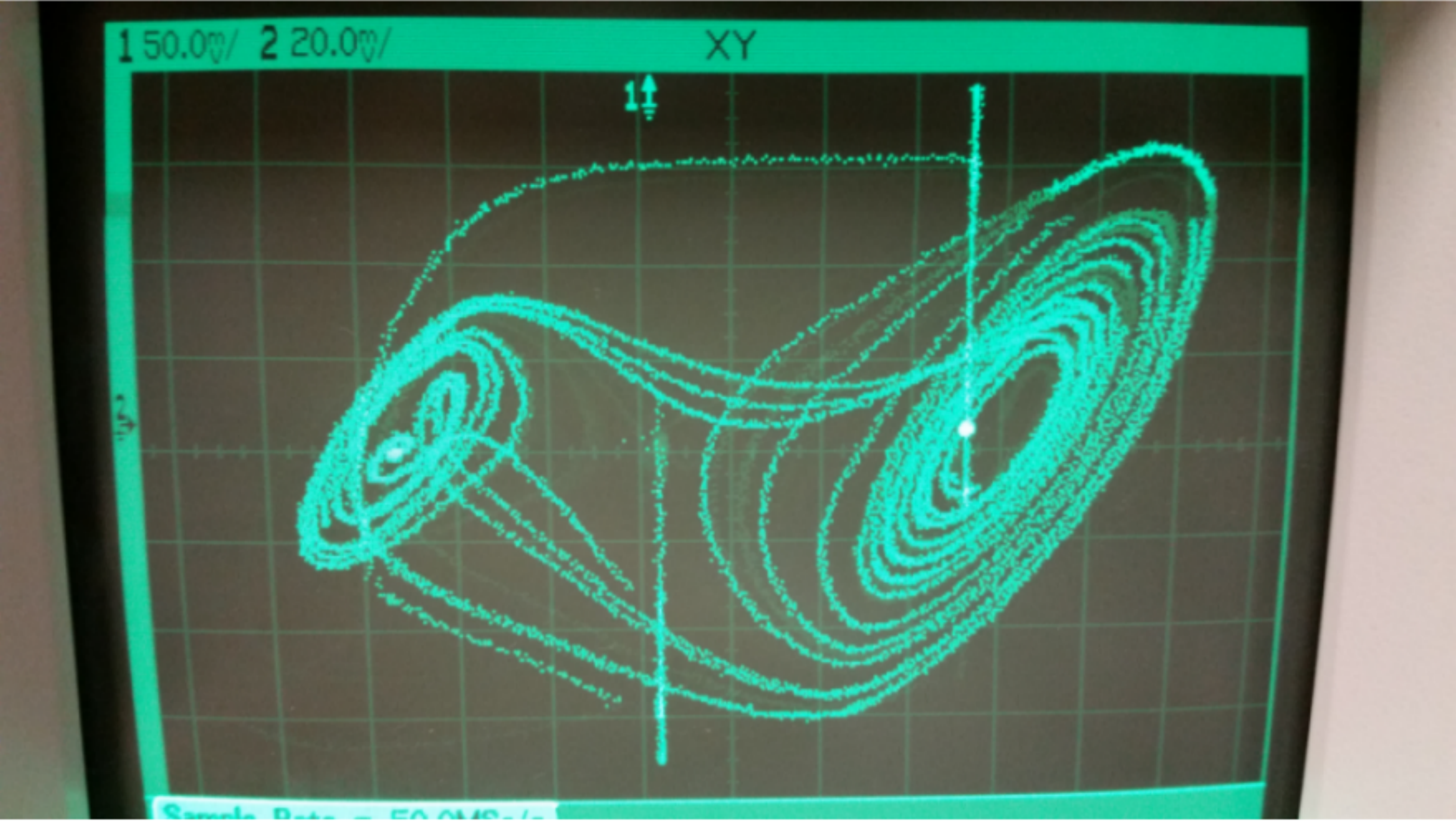}
\caption{Capacitor voltages}\label{Fig: Exp_Double_Scroll}
\end{subfigure}
\begin{subfigure}{0.49\textwidth}
\includegraphics[width = \textwidth]{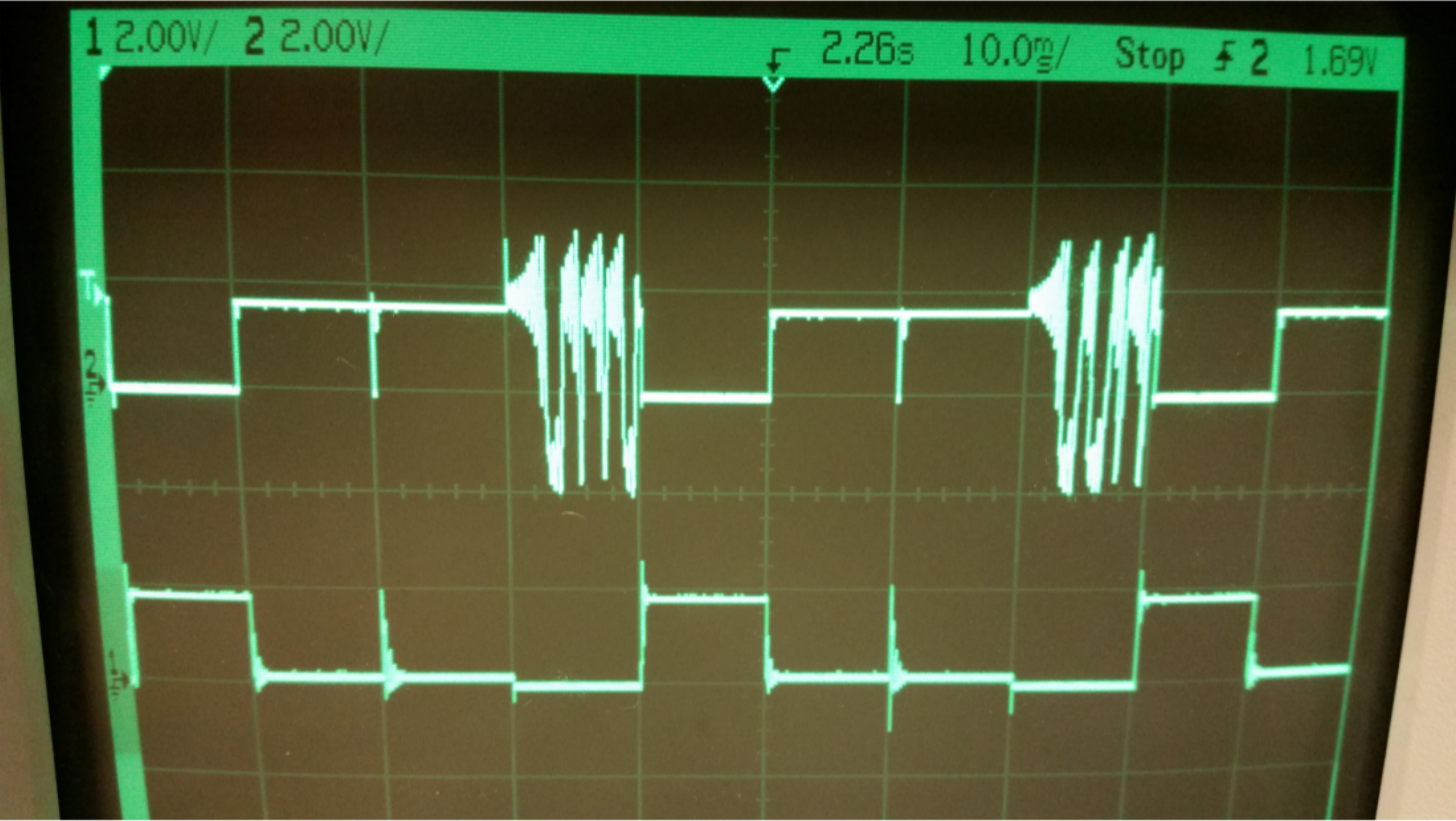}
\caption{Threshold 1 and NOR output 1}\label{Fig: Exp_threshold1}
\end{subfigure}
\begin{subfigure}{0.49\textwidth}
\includegraphics[width = \textwidth]{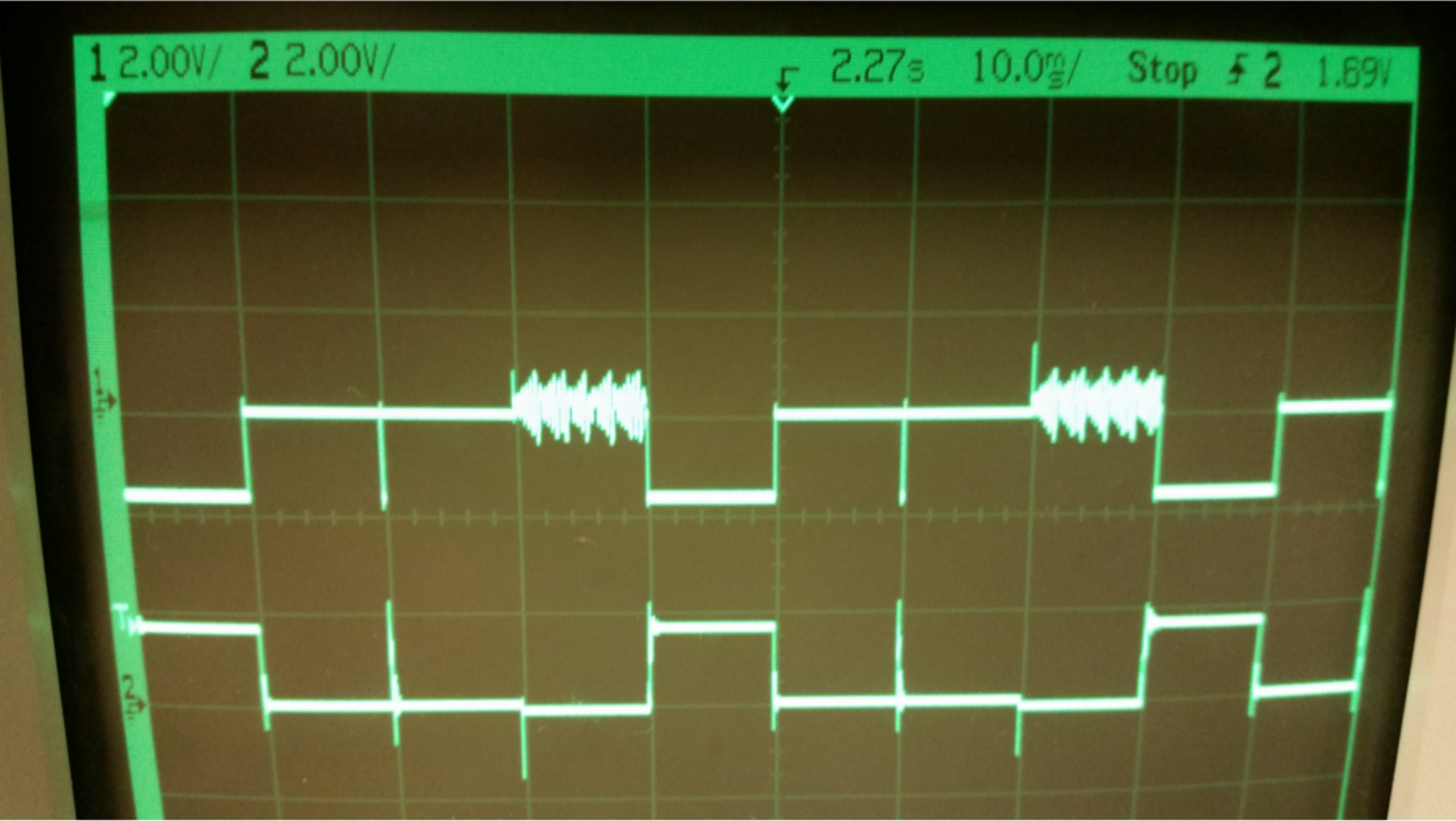}
\caption{Threshold 2 and NOR output 2}\label{Fig: Exp_threshold2}
\end{subfigure}
\caption{Experimental results for the dual NOR gate setup.
For (a), (c), and (d) the abscissa represents time and the ordinate represents voltage.
In (b) the plot shows the phase space produced by the two capacitors.  In (a) the plot shows
the two input voltages.  In (c) and (d) the plot shows the respective threshold voltage and NOR
outputs.}\label{Fig: NOR_ExpResults}
\end{figure}

Since the circuit of Cafagna and Grassi \cite{Cafagna-Grassi06} uses components that have become
unavailable, we needed to design a modern chaotic RSFF/dual NOR gate.
The schematic of our circuit design is shown in Fig. \ref{Fig: RSFFschem}.  For the \emph{MultiSIM} simulations
we use the components shown in Table \ref{Tab: Parts}.  The chaotic backbone of the circuit comes
from Chua's circuit, and we employ threshold control units to produce the logical outputs.  The \emph{MultiSIM}
plots for the dual NOR gate are shown in Fig. \ref{Fig: Ian_NOR}.
We then made a physical realization of the circuit with mostly the same components (Table \ref{Tab: Parts})
as the \emph{MultiSIM} simulations.
The experimental setup is shown in Fig. \ref{Fig: ExpSetup}.  In order to power the circuit and get readings
we used a DC power supply, wave form generator, \emph{Arduino Due}\textregistered, and an oscilloscope.
From the physical realization we reproduced the \emph{MultiSIM} plots.  To activate the dual NOR gate we
keep switches 1 and 2 in Fig. \ref{Fig: RSFFschem} closed and switches 3 and 4 open.  The experimental results for the
dual NOR gate setup are shown in Fig. \ref{Fig: NOR_ExpResults}.  In Figs. \ref{Fig: Exp_threshold1}
and \ref{Fig: Exp_threshold2} we notice windows of what looks like chaotic dynamics.  This is also observed
in the \emph{MultiSIM} simulations.  It should be noted that the threshold behavior of the \emph{MultiSIM} circuit
and the physical realization are slightly different due to a change in op-amps.
We take a detailed look at the dynamics in Sec. \ref{Sec: Deterministic}.

\begin{figure}[htbp]
\begin{subfigure}[t]{0.32\textwidth}
\includegraphics[width = \textwidth]{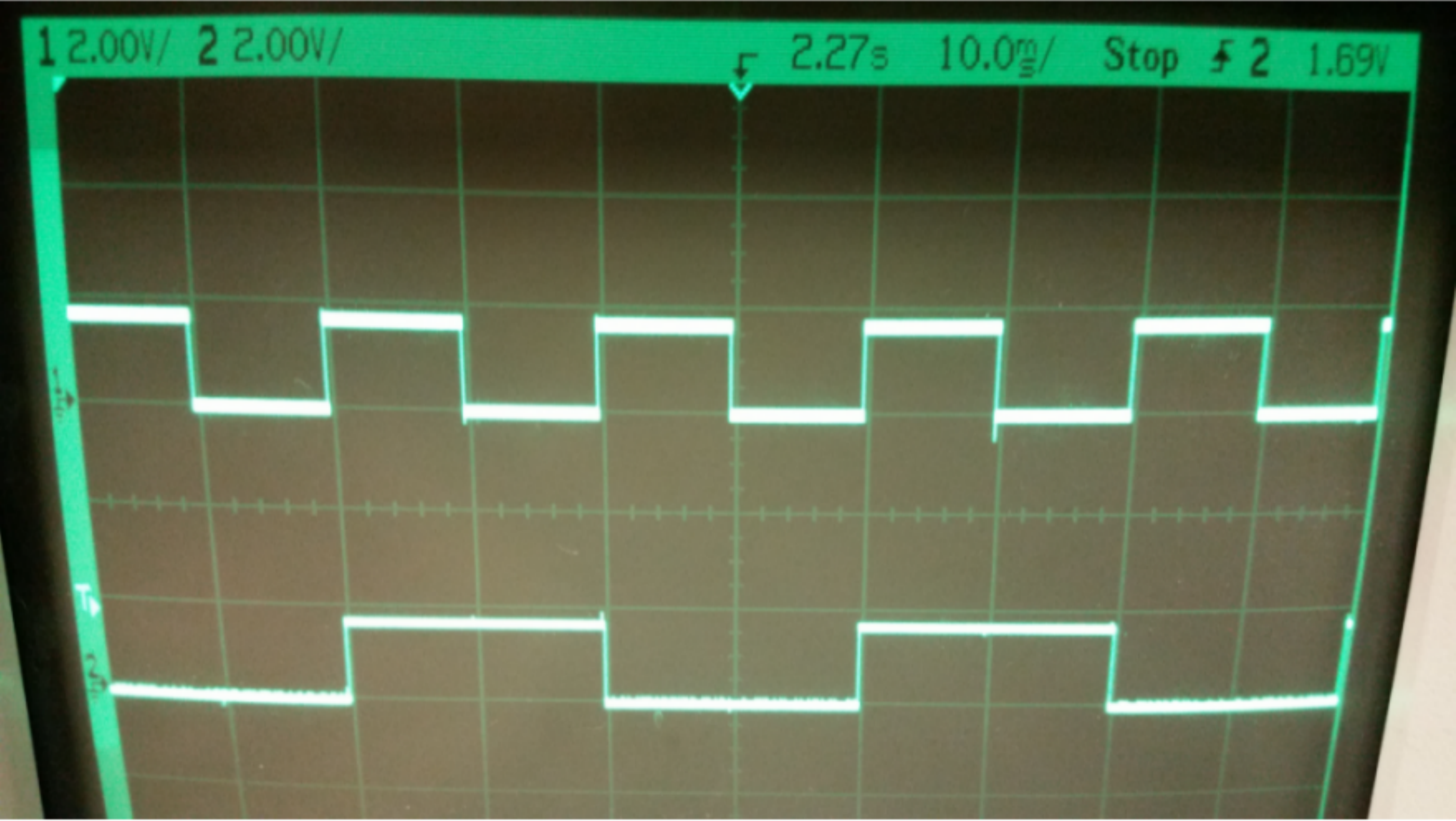}
\caption{Inputs}
\end{subfigure}
\begin{subfigure}[t]{0.32\textwidth}
\includegraphics[width = \textwidth]{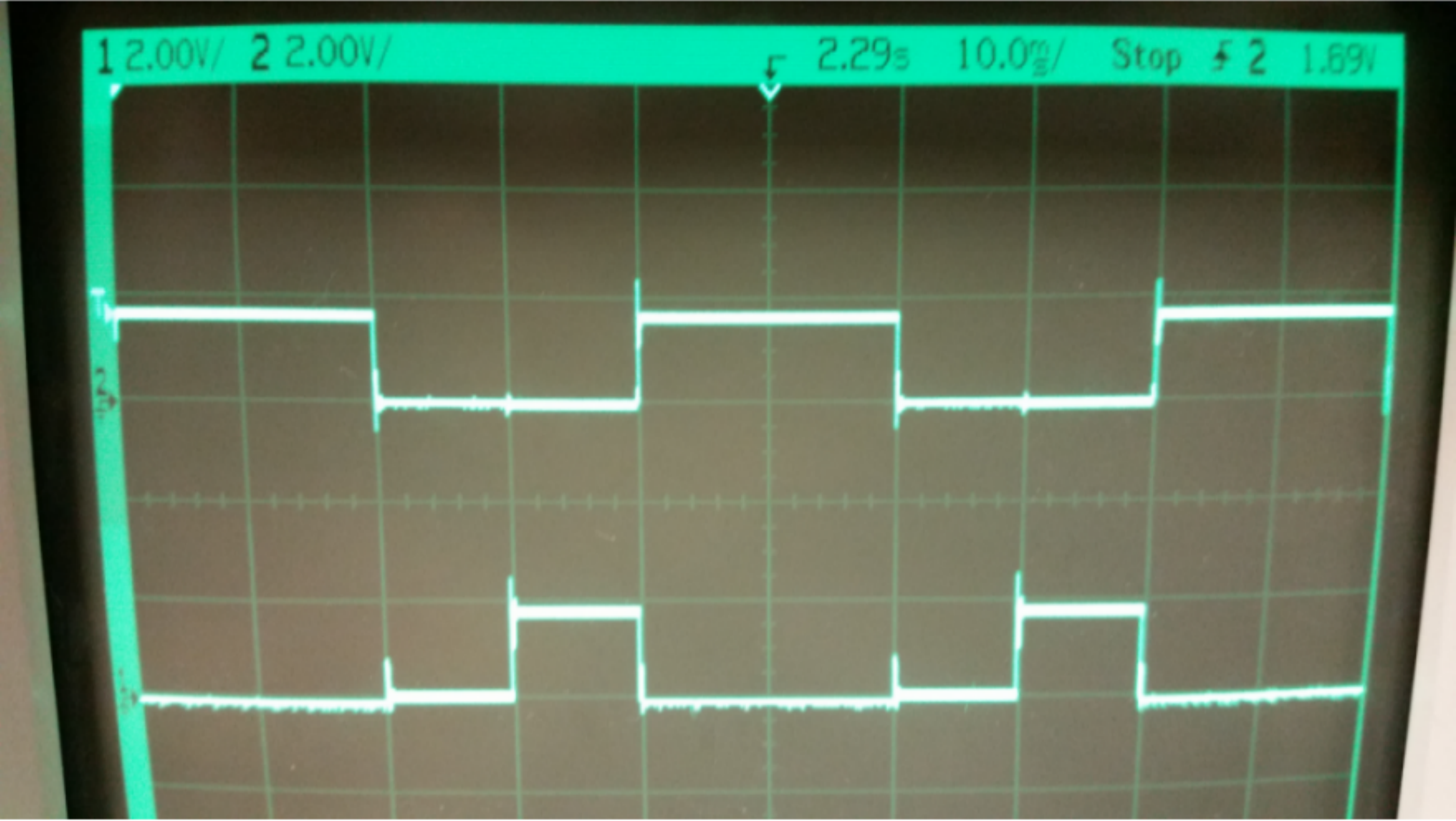}
\caption{Output without explicit feedback}\label{Fig: RSFF_Output}
\end{subfigure}
\begin{subfigure}[t]{0.32\textwidth}
\includegraphics[width = \textwidth]{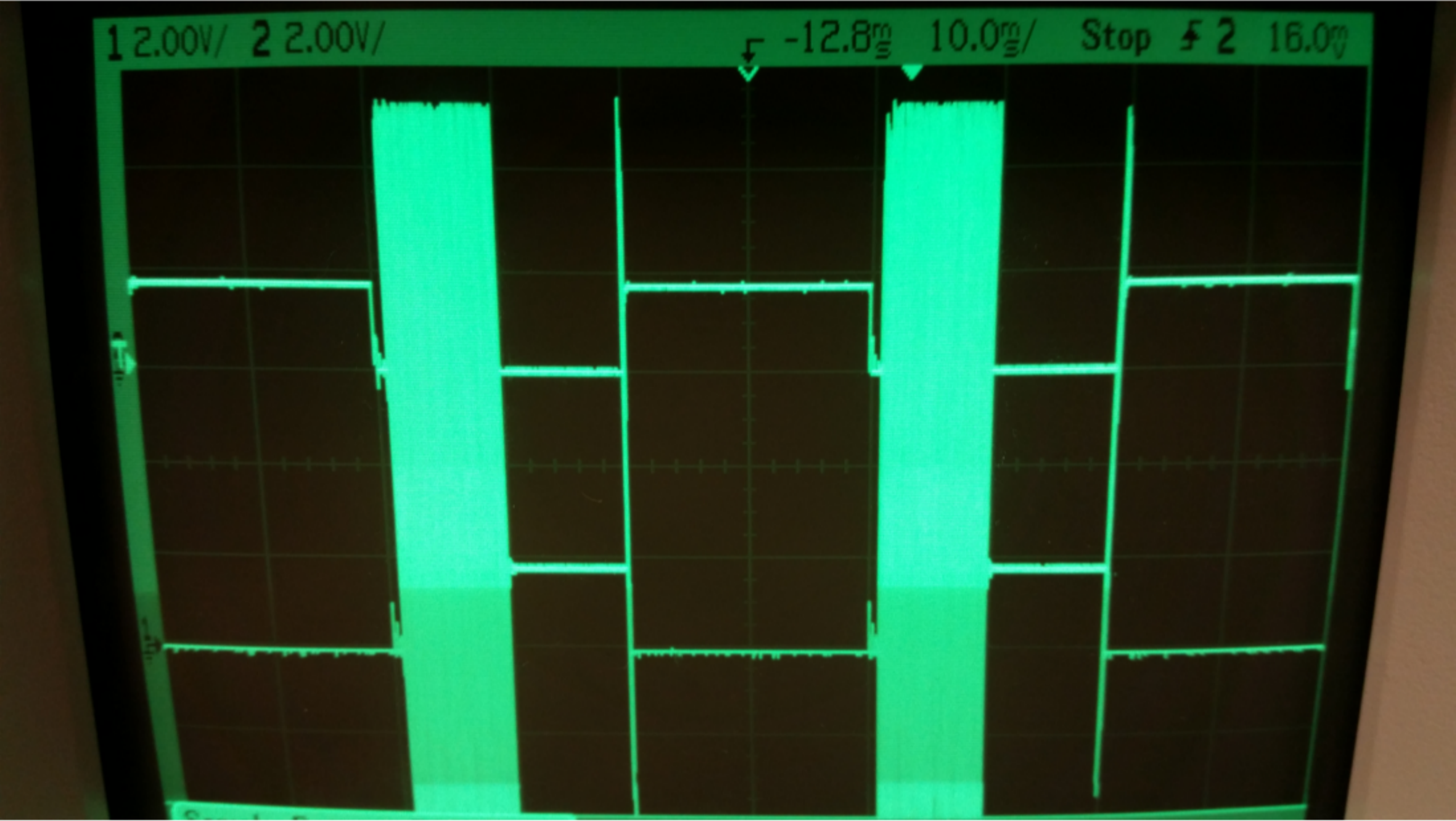}
\caption{Output with explicit feedback}
\end{subfigure}
\caption{Experimental results for the RSFF setup.
For (a), (b), and (c) the abscissa represents time and the ordinate represents voltage.
In (a) the plot shows the two input voltages.  In (b) the voltage was measured when
switches 3 and 4 in Fig. \ref{Fig: RSFFschem} are open.  In (c) the voltage was measured
when switches 3 and 4 in Fig. \ref{Fig: RSFFschem} are closed.}
\label{Fig: RSFF_ExpResults}
\end{figure}

To deactivate the dual NOR gate we open switches 1 and 2 in Fig. \ref{Fig: RSFFschem}.  To properly activate the RSFF
we need to close switches 3 and 4, however we get the RSFF outputs even with the
switches open.  When we close the switches we notice the race conditions for the ``1-1''
input causing wild oscillations, which is completely missed in the simulations.
This shows that the design has implicit feedbacks, through
the op-amps, which results in RSFF operations.  However, when the two NOR gates are
explicitly connected with one another (i.e. the output of one NOR gate feeds back to the
input of the other NOR gate) the race condition exacerbates the oscillations.
The experimental results for the RSFF setup are shown in Fig. \ref{Fig: RSFF_ExpResults}.
While the experiments of the RSFF mainly work as expected, the \emph{MultiSIM} simulations are
not able to properly capture this behavior.

\section{Deterministic models}\label{Sec: Deterministic}

\begin{figure}[htbp]
\centering
\begin{subfigure}{0.45\textwidth}
\centering
\includegraphics[width = \textwidth]{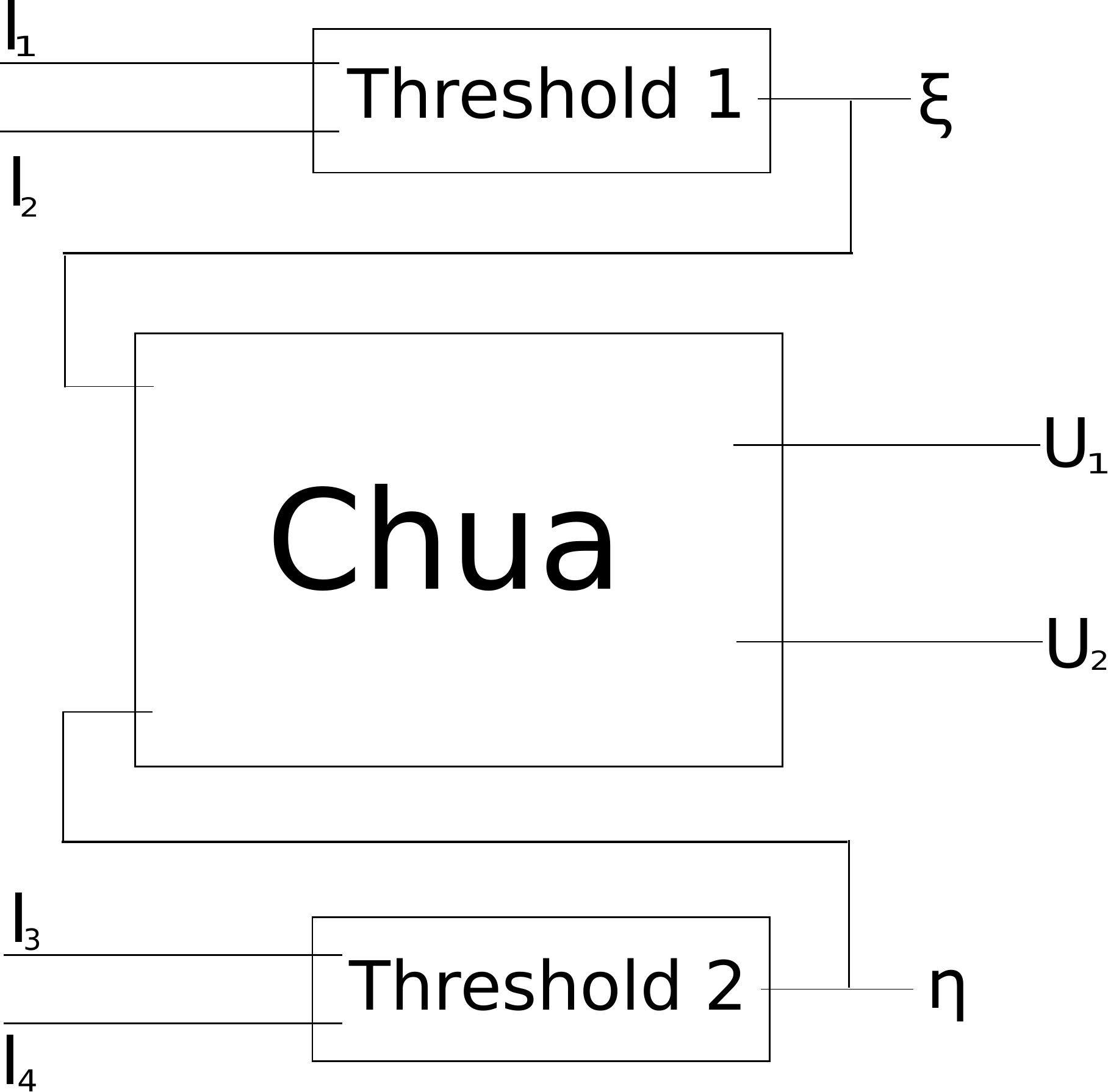}
\caption{Simplified schematic of two chaotic NOR gates via Chua's circuit and TCUs}\label{Fig: NORDiag}
\end{subfigure}\qquad
\begin{subfigure}{0.45\textwidth}
\centering
\includegraphics[width = \textwidth]{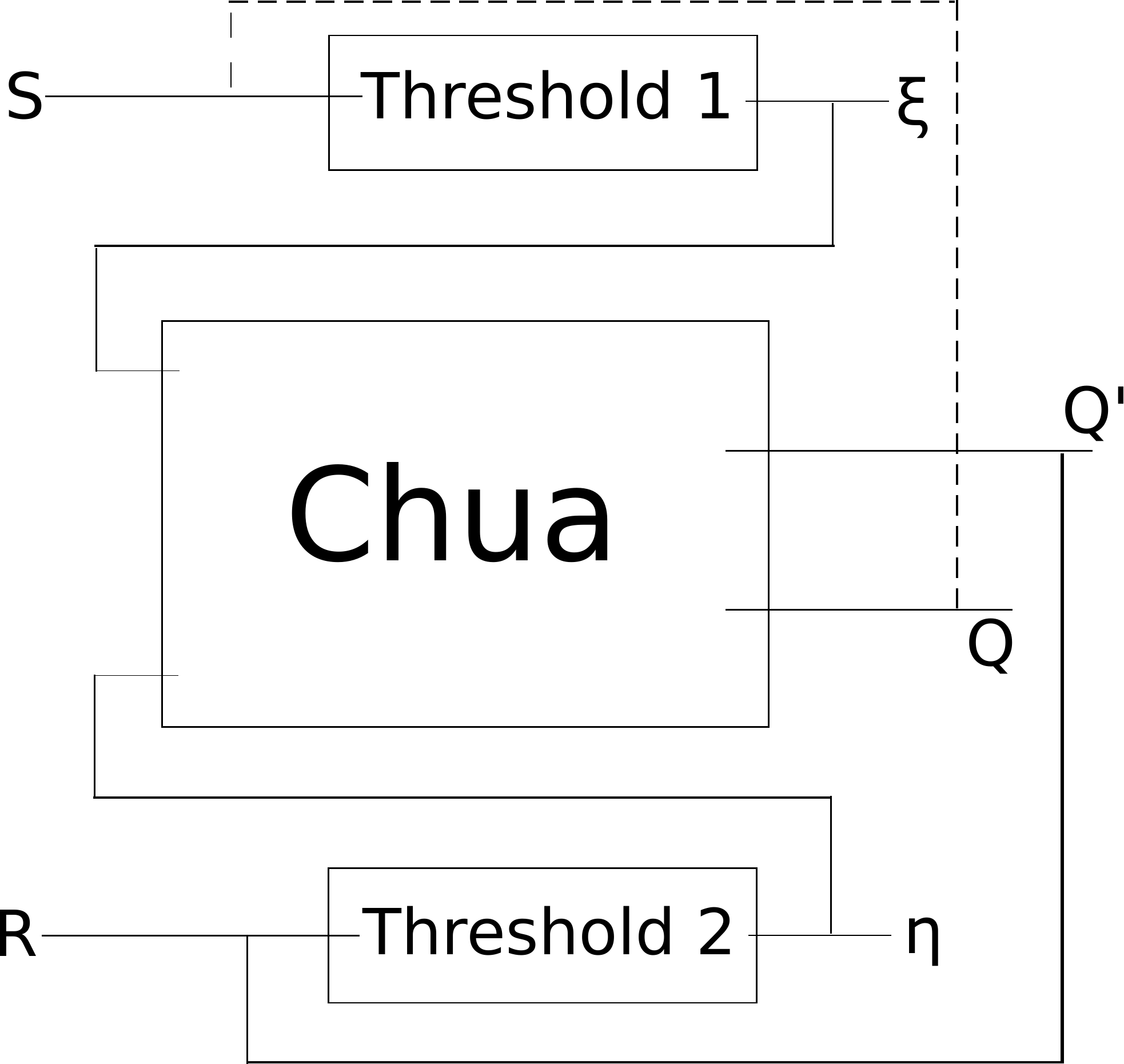}
\caption{Simplified schematic of a chaotic RS flip-flop via Chua's circuit and TCUs}\label{Fig: RSFFDiag}
\end{subfigure}
\caption{Simplified schematics used to illustrate the building blocks of the models.}\label{Fig: Diags}
\end{figure}

We first model the NOR gates, then we use those models to model the RSFF.
This is accomplished by constructing continuous extensions and approximating
the behavior of the TCU and Chua's circuit.  We have also provided simplified
schematics in Fig. \ref{Fig: Diags} to help illustrate the modeling process.
It should be noted, that while in the experiments we use a high voltage of
$1.84$ volts, in our models this is normalized to $1$.

One may suggest
solving Matsumoto's equation \cite{Matsumoto84} for Chua's circuit as part
of the model.  While
this would be a legitimate approach, the goal of the article is to formulate the
simplest model in terms of derivation and simulation.  Having to solve the ordinary
differential equation at each time step would be reasonable for a two gate
circuit, but would not be scalable to large chaotic logical circuits with many
gates.  For these reasons, we forgo explicitly modeling the capacitor voltages, and instead
develop models for the threshold voltages and the outputs, then show that the
derived capacitor voltages agree reasonably well with experiments and \emph{MultiSIM} simulations in addition to
showing the threshold and outputs agree surprisingly well with the experiments and simulations.

\subsection{NOR gate}

Here we derive and analyze the NOR gate models.  The schematic for the gates
given in Fig. \ref{Fig: RSFFschem} may be simplified to Fig. \ref{Fig: NORDiag}
in order to get a general picture of the circuit.  It should be noted that the blocks
are not absolutely accurate as the block labeled ``Chua'' has additional op-amps
to that of the traditional Chua circuit and feedback from the ``Chua'' block to the
threshold blocks are not explicitly shown.

\subsubsection{Derivation}

We begin by formulating the simplest continuous extension of the NOR operator:
$x \nor y = (1-x)(1-y)$.  This gives us the equation,
\begin{equation}\label{Eq: NOR}
\begin{split}
U_1 &= (1-I_1)(1-I_2),\\
U_2 &= (1-I_3)(1-I_4);
\end{split}
\end{equation}
where $U$ are the respective outputs and $I$ are the respective inputs as shown
in Fig. \ref{Fig: NORDiag}.  This na\"{i}ve extension is enough to approximate
the behavior for the NOR gate outputs, and we shall also modify them later to simulate
certain subtle aspects of the circuit.  However, the more interesting dynamics
are observed in the TCUs.

First, let us define the maps, $f: [0,1]^2\times[-2,2] \rightarrow [-2,2]$ and
$g: [0,1]^2\times[-1,1] \rightarrow [-1,1]$, where the chosen intervals are
the operating domains of the TCUs by design.  We apply these maps to the
inputs and current time threshold voltages to get the next time threshold
voltages,
\begin{equation}\label{Eq: Thresh}
\begin{split}
\xi_{n+1} &= f(I_1,I_2,\xi_n),\\
\eta_{n+1} &= g(I_1,I_2,\eta_n);
\end{split}
\end{equation}
where $\xi$ and $\eta$ are the respective threshold voltages as shown in
Fig. \ref{Fig: NORDiag}.

Next, we tabulate our observations from Figs. \ref{Fig: NOR_ExpResults} and \ref{Fig: Ian_NOR} in
(\ref{Eq: Tab}), and formulate functions $f$ and $g$ that qualitatively
reproduce the observed behavior.  It should be
noted that henceforth we shall denote a seemingly random voltage
(in the chaotic regime) by a star ($\star$), where $\star$ is not a number
to be measured, but rather denotes chaotic behavior.

\begin{equation}\label{Eq: Tab}
\begin{array}{ccc}
f(0,0,0) = 0 & \qquad f(1,0,1) = f(0,1,1) = 1 \qquad & f(1,1,\star) = \star,\\
g(0,0,-1) = -1 & \qquad g(1,0,0) = g(0,1,0) = 0 \qquad & g(1,1,\star) = \star.
\end{array}
\end{equation}

In order to satisfy these criteria, we first define the maps, $y_f: [-2,2] \rightarrow [-2,2]$
and $y_g: [-1,1] \rightarrow [-1,1]$, then
\begin{equation}\label{Eq: ThreshFcns}
\begin{split}
f(I_1,I_2,x_1) &:= |I_1 - I_2| + I_1I_2y_f(x_1),\\
g(I_3,I_4,x_2) &:= |I_3 - I_4| - 1 + I_3I_4y_g(x_2);
\end{split}
\end{equation}
satisfies each property except chaotic behavior.

\begin{figure}
\begin{subfigure}[t]{0.44\textwidth}
\includegraphics[width = \textwidth]{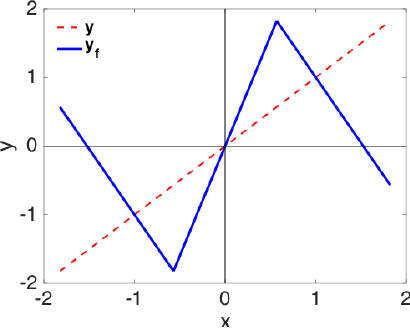}
\caption{Illustration of $y_f$.}\label{Fig: y_f}
\end{subfigure}\qquad
\begin{subfigure}[t]{0.46\textwidth}
\includegraphics[width = \textwidth]{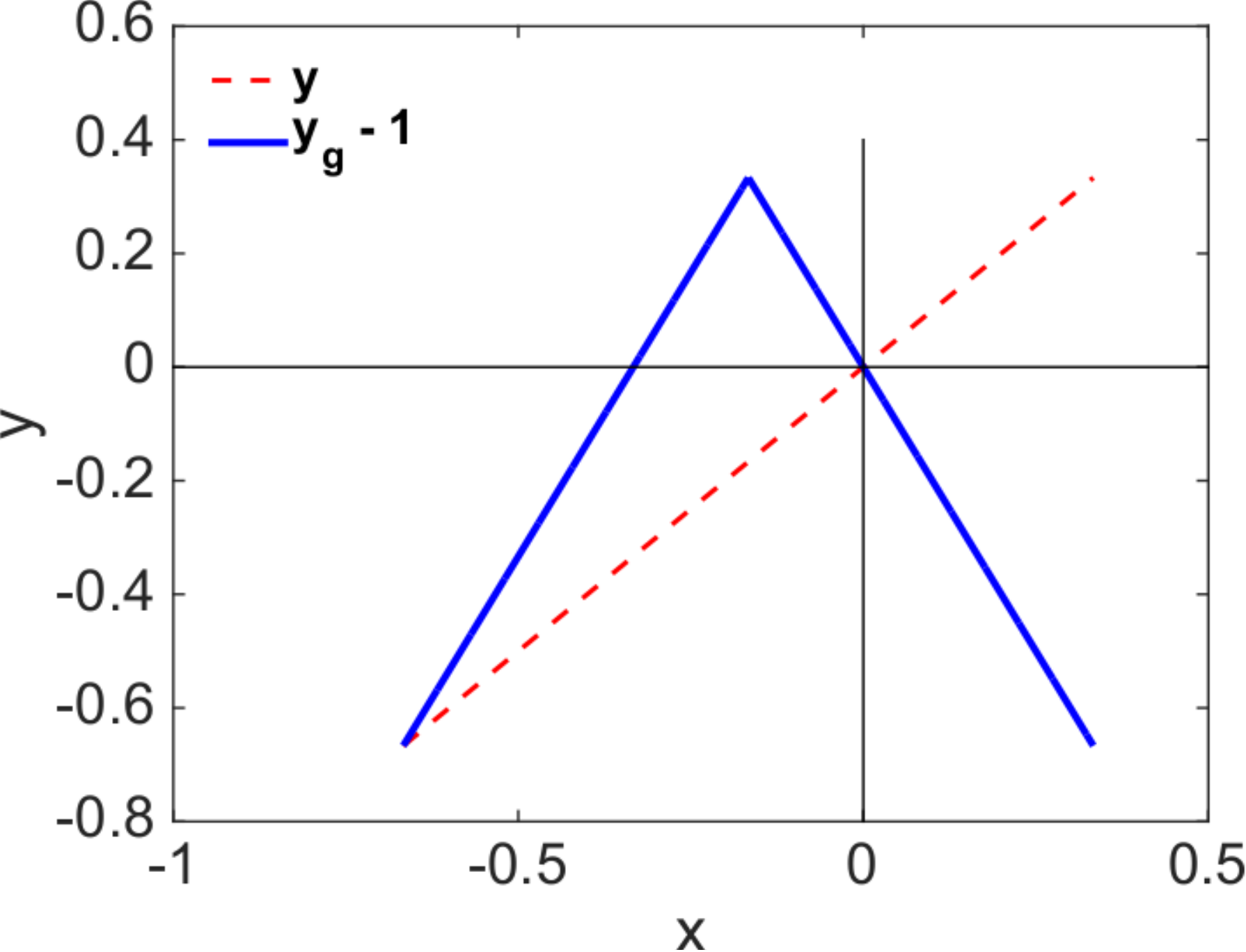}
\caption{Illustration of $y_g - 1$.}\label{Fig: y_g}
\end{subfigure}
\caption{Plots of $y_f$ and $y_g$ for fixed values of $\mu_f$, $\nu_f$, $\mu_g$, $\nu_g$.}
\label{Fig: ThreshFcns}
\end{figure}

Now, we must formulate $y_f$ and $y_g$ such that (\ref{Eq: ThreshFcns})
reproduces the qualitative behavior of the threshold voltages for $(I_1,I_2) = (1,1)$
and $(I_3,I_4) = (1,1)$ as shown in Fig. \ref{Fig: NOR_ExpResults}.  We postulate
$y_f$ and $y_g$ will be tent map - like, which is reasonable in the physical
sense since these op-amps influence the signal in an approximately piecewise
linear manner.  Furthermore, the NOR gate developed by Murali \ea employed
a TCU designed to behave in a similar manner to that of a logistic map \cite{MSD2003A}, and a
tent map is topologically conjugate to a logistic map.  For $y_f$ we write,
\begin{equation}\label{Eq: y_f}
y_f(x) := \begin{cases}
\frac{1+\mu_f}{1-\mu_f}(x+\nu_f) - \mu_f\nu_f  &x \in [-\mu_f\nu_f,-\nu_f],\\
\qquad \mu_f x & x \in [-\nu_f,\nu_f],\\
\frac{1+\mu_f}{1-\mu_f}(x-\nu_f) + \mu_f\nu_f  &x \in [\nu_f,\mu_f\nu_f];
\end{cases}
\end{equation}
and we write $y_g$ as,
\begin{equation}\label{Eq: y_g}
y_g(x) := \nu_g + \mu_g \begin{cases}
1+x-\nu_g & \nu_g - 1 \leq x \leq \nu_g - 1/2,\\
\nu_g - x & \nu_g - 1/2 \leq x \leq \nu_g .
\end{cases}
\end{equation}
We illustrate (\ref{Eq: y_f}) and (\ref{Eq: y_g}) in Fig. \ref{Fig: ThreshFcns}.
In Fig. \ref{Fig: y_f} and (\ref{Eq: y_f}), $\mu_f$ is the slope of the middle line
segment and it also determines the slope of the other two line segments, whereas
$\nu_f$ mainly affects the domain of the map.
In Fig. \ref{Fig: y_g} and (\ref{Eq: y_g}), $\mu_g$ is twice the height of the
map and $1-\nu_g$ is the amount the new interval is translated from the interval $[0,1]$.
Moreover, all parameters are positive.

This gives us a model of the thresholding in the two NOR gate operation of the circuit.

\subsubsection{Basic properties}

Since the interesting behavior arises for $(I_1,I_2) = (1,1)$ and $(I_3,I_4) = (1,1)$,
it suffices to analyze $f(1,1,x)$ and $g(1,1,x)$.  First we search for the fixed points.
From empirical observations we require $g(1,1,x)$ and $f(1,1,x)$ to have a fixed point at the
origin and $f(1,1,x)$ to have two other nonzero fixed points.  By definition, the latter
has a fixed point at the origin and two nonzero fixed points.  However, for the
former, we require the right branch to intersect the origin.  In order to
achieve this, we set $g(1,1,x) = x = 0$ for $x \in [\nu_g-1/2,\nu_g]$(the right
branch),
\begin{equation*}
0 = -1 + \nu_g + \mu_g\nu_g \Rightarrow \nu_g = \frac{1}{1+\mu_g}.
\end{equation*}
Furthermore, in order to ensure the inequality, $\nu_g - 1/2 \leq x \leq \nu_g$,
we require $\nu_g \leq 1/2$, i.e. $\mu_g \geq 1$.

Now let us identify any additional fixed points for $g(1,1,x)$, which necessarily
lies on the left branch, i.e. $x \in [\nu_g - 1,\nu_g - 1/2]$,
\begin{align*}
x_* &= -1 + \nu_g + \mu_g - \nu_g\mu_g + \mu_g x_*\\
\Rightarrow (1-\mu_g)x_* &= -1 + \frac{1}{1+\mu_g} + \mu_g - \frac{\mu_g}{1+\mu_g}
= -(1-\mu_g) + \frac{1-\mu_g}{1+\mu_g}\\
\Rightarrow x_* &= -1 + \frac{1}{1+\mu_g} = -\frac{\mu_g}{1+\mu_g}.
\end{align*}

Next, we identify the two nonzero fixed points for $f(1,1,x)$.  For $x \in [\pm\nu_f,\pm\nu_f\mu_f]$,
\begin{equation*}
x_* = \frac{1+\mu_f}{1-\mu_f}x_* \pm \left(\mu_f\nu_f - \nu_f\frac{1+\mu_f}{1-\mu_f}\right)
\Rightarrow \left(\frac{-2\mu_f}{1-\mu_f}\right)x_* = \mp \nu_f\frac{1+\mu_f^2}{1-\mu_f}
\Rightarrow x_* = \pm \nu_f\frac{1+\mu_f^2}{2\mu_f}.
\end{equation*}
For the sake of convenience, we outline the fixed points in Table \ref{Tab: NOR_fp}.

To analyze the stability of the fixed points of $g(1,1,x)$, let us take the derivative,
\begin{equation}
\frac{d g(1,1,x)}{dx} = \begin{cases}
\mu_g & \nu_g - 1 \leq x \leq \nu_g - 1/2,\\
-\mu_g & \nu_g - 1/2 \leq x \leq \nu_g ;
\end{cases}
\end{equation}
Both fixed points are sources when $\mu_g > 1$.  For $\mu_g = 1$, there
is a line of fixed points on the left branch such that it constitutes a global
attracting set.  Clearly, this case is unphysical, and hence we require $\mu_g > 1$.

Taking the derivative for $f(1,1,x)$ gives,
\begin{equation}
\frac{d f(1,1,x)}{dx} = \begin{cases}
\frac{1+\mu_f}{1-\mu_f} & x \in [-\mu_f\nu_f,-\nu_f],\\
\:\mu_f & x \in [-\nu_f,\nu_f],\\
\frac{1+\mu_f}{1-\mu_f} & x \in [\nu_f, \mu_f\nu_f];
\end{cases}
\end{equation}
By definition, $\mu_f \geq 1$, since otherwise the two outer branches would be undefined.
If $\mu_f = 1$, there is a line of fixed points on the middle branch, and hence this too is
unphysical.  Therefore, we require $\mu_f > 1$, which shows the origin is a source.  For
the other two fixed points, if $(1+\mu_f)/(1-\mu_f) < -1$ they are be sinks and if
$(1+\mu_f)/(1-\mu_f) > -1$ they are sources.  Notice that $(1+\mu_f)/(1-\mu_f) = -1$
only when $\mu = 0$, which we showed was impossible.  Now, if $(1+\mu_f)/(1-\mu_f) < -1$,
$1 < -1$, which is false.  If $(1+\mu_f)/(1-\mu_f) > -1$, $1 > -1$, which is true.  Therefore,
these fixed points are always sources.

The fixed points along with the conditions on the parameters required to yield
physical results are summarized in the table below,
\begin{table}[htbp]
\centering
\setlength{\tabcolsep}{24pt} 
\renewcommand{\arraystretch}{3}
\begin{tabular}{c||c|c}
Equation & Fixed points & Conditions\\
\hline
$g(1,1,x)$ & $x_* = 0,\, -\dfrac{\mu_g}{1+\mu_g}$ & $\mu_g > 1$\\
$f(1,1,x)$ & $x_* = 0,\, \pm \nu_f\frac{1+\mu_f^2}{2\mu_f}$ & $\mu_f > 1$ and $\nu_f > 0$\\
\end{tabular}
\caption{Summary of fixed points and conditions on parameters to cause all fix points to be sources.}
\label{Tab: NOR_fp}
\end{table}

\subsubsection{Chaos}

Here we shall prove the maps $g(1,1,x)$ and $f(1,1,x)$ become chaotic for certain parameters.
First we prove this for $g(1,1,x)$, which employs a simple translation to the tent map.

\begin{thm}
The map $g(1,1,x)$ is chaotic for $\mu_g \geq 2$.  In addition, it has a nonwandering set
in the form of a translated Cantor set on $[1-\nu_g,\nu_g]$ for $\mu_g > 2$.  Moreover,
for $\mu = 3$, the nonwandering set is the middle-third Cantor set translated to the
interval $[1-\nu_g,\nu_g]$.

\begin{proof}
Consider the translation $(u,v) = H(x,y):\: [1-\nu_g,\nu_g]^2 \rightarrow [0,1]^2$,
defined as $(u, v) = (x + 1 - \nu_g, g(1,1,x) + 1 - \nu_g)$, applied to $g(1,1,x)$.
This produces the map $v = T_{\mu_g(u)}:\: [0,1] \rightarrow [0,1]$, which is exactly the
tent map.  Since $H$ is a homeomorphism, it suffices to analyze the tent map.  It
is well known that
when $\mu \geq 2$, the tent map is chaotic.  Furthermore, for $\mu > 2$, the nonwandering set
is a Cantor set, and for $\mu = 3$, it is precisely the middle-third Cantor set.  This shows the
map $g(1,1,x)$ is also chaotic for $\mu_g \geq 2$, and has a nonwandering set in the
form of a translated Cantor set, thereby completing the proof.
\end{proof}

\end{thm}

Now we prove $f(1,1,x)$ is chaotic in the physical parameter regime outlined in Table \ref{Tab: NOR_fp}.
For the sake of brevity, we assume the parameters are in this regime for our next theorem.
The main idea of the proof is to search for 3-cycles and use the main theorem by Li
and Yorke \cite{LiYorke75}.  Since the formula for $f^3$ becomes overly complex, we shall
use properties of $f^3$ to show the existence of a 3-cycle rather than finding it explicitly,
and we provide visual aids to illustrate the proof.

\begin{figure}
\begin{subfigure}[t]{0.45\textwidth}
\includegraphics[width = \textwidth]{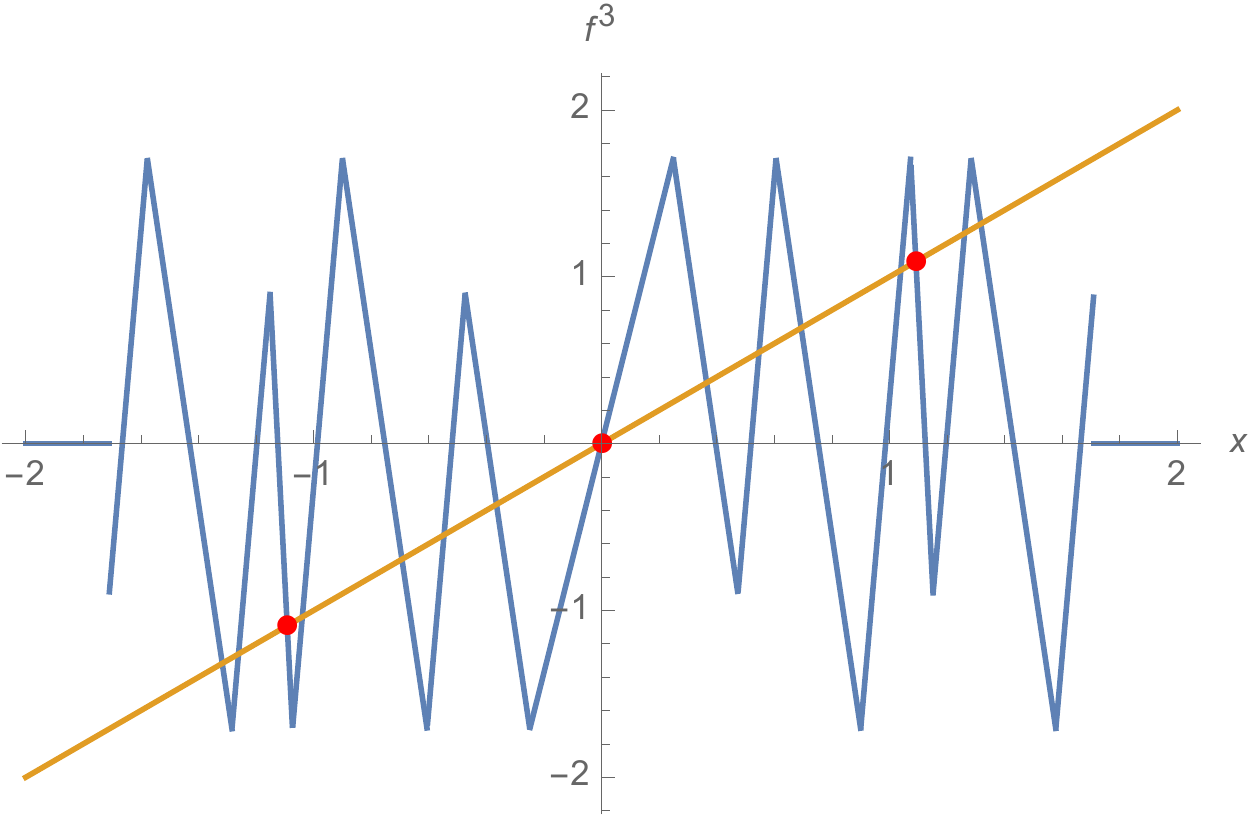}
\caption{$\mu_f = 1.9$, $\nu_f = 0.9$}\label{Fig: 3CycleA}
\end{subfigure}
\begin{subfigure}[t]{0.45\textwidth}
\includegraphics[width = \textwidth]{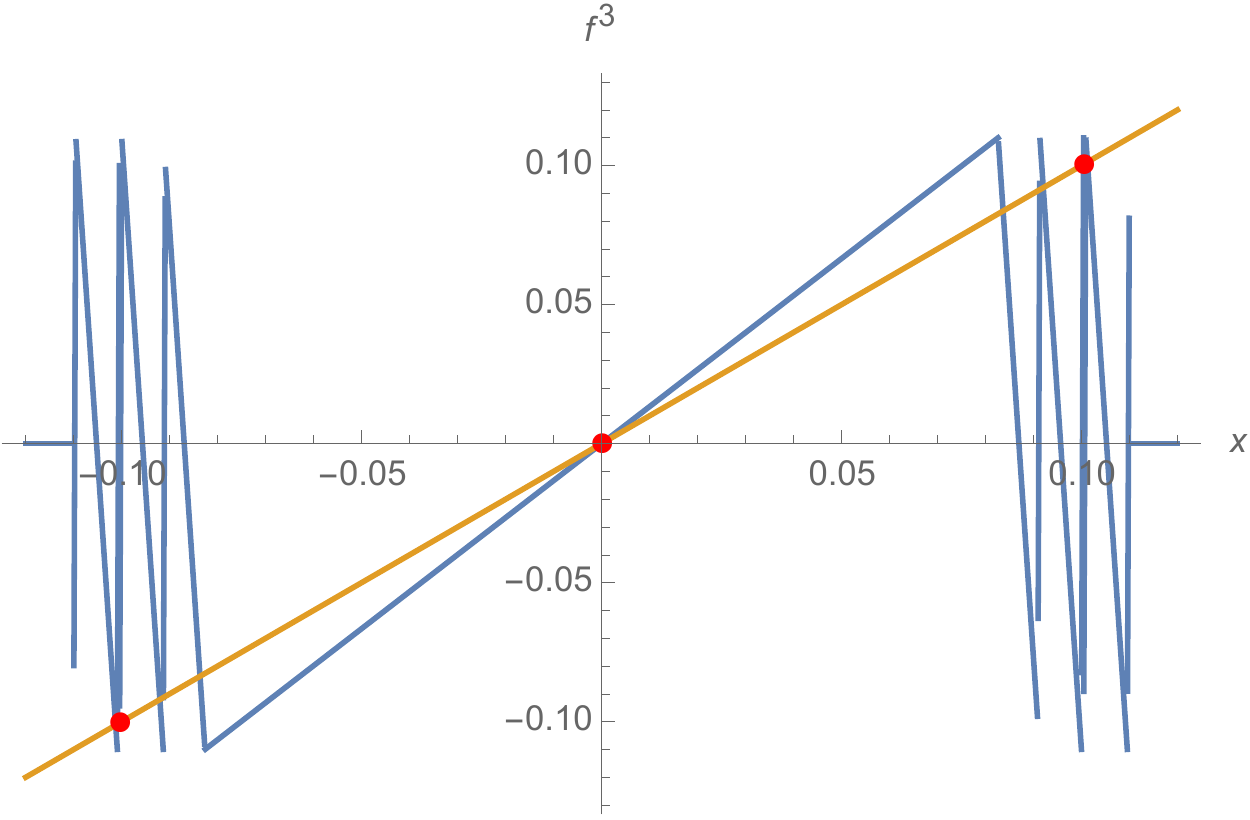}
\caption{$\mu_f = 1.1$, $\nu_f = 0.1$}\label{Fig: 3CycleB}
\end{subfigure}
\caption{Plots of $f^3$ for parameters in the physical regime of Table \ref{Tab: NOR_fp}
with parameters similar to that of the experiments and simulation and parameters just
within the physical regime respectively.  It should be noted that the gaps between
certain branches in Fig. \ref{Fig: 3CycleB} are due to computational inaccuracy and in reality
all branches connect.}
\label{Fig: 3Cycle}
\end{figure}

\begin{thm}
For every $n \in \N$ there exists a periodic point $p_n \in [-\mu_f\nu_f,\mu_f\nu_f]$
of the map $f(1,1,x)$ having period $n$ and $[-\mu_f\nu_f,\mu_f\nu_f]$ contains
chaotic orbits of $f$.  Furthermore, there exists an uncountable set $S \subset [-\mu_f\nu_f,\mu_f\nu_f]$
containing no periodic orbits.

\begin{proof}
It is shown in \cite{LiYorke75} that a 3-cycle implies chaos.  Now, we show there exists
a 3-cycle for all parameter values in the physical regime.  This is done by finding the roots
of $f^3$ not including the fixed points, or rather showing they exist.

First, let $P_n$ be the set of roots of $f^n$, then $P_3 \setminus P_1$ is the set of points
having period $3$.  We observe $f^3$ (Fig. \ref{Fig: 3Cycle}) has $17$ linear branches.
Since these are linear, each branch may intersect the line $y=x$ at most once.  The two
outermost branches will never intersect the line $y=x$ for the parameter regime outlined
in Table \ref{Tab: NOR_fp}.  Then $\max(\card(P_3)) = 15$ and $\card(P_1) = 3$ (i.e. 
$f$ has three fixed points), hence $\max(\card(P_3 \setminus P_1)) = 12$.  Notice,
a 3-cycle exists only if $\card(P_3 \setminus P_1) \in \{6,12\}$ due to the symmetry.

We observe (Fig. \ref{Fig: 3CycleA}) $\card(P_3 \setminus P_1) = 12$ for the parameters
near that used in simulations.  If $\mu_f$ and $\nu_f$ are varied forward we maintain
$\card(P_3 \setminus P_1) = 12$.  If we vary them backward, the first instance
$\card(P_3 \setminus P_1) \neq 12$ occurs when the cusps of the second and third
branches from the left, and respectively from the right, lie on the line $f(1,1,x) = x$.
The cusps are located at
\begin{equation*}
\hat{x} = \pm\left(\nu_f + \mu_f\nu_f\frac{\mu_f - 1}{\mu_f + 1}\right.
\left. - \frac{\nu_f}{\mu_f}\frac{\mu_f - 1}{\mu_f + 1}\right),\,
f(1,1,\hat{x}) = \pm\mu_f\nu_f.
\end{equation*}
Then, solving $\hat{x} = f(1,1,\hat{x})$ for $\mu_f$, gives $\mu_f = 1$, which is unphysical.
Since $\card(P_3 \setminus P_1) = 12$ in the parameter regime of Table \ref{Tab: NOR_fp},
3-cycles (in fact, four of them) exist.  Therefore, the hypotheses of \cite{LiYorke75} is
satisfied, thereby completing the proof.
\end{proof}

\end{thm}

\subsection{Set/reset flip-flop}

Now that we have a model for the NOR gates we can derive the model of the RSFF.
Just as with the NOR gate, we sketch a simplified schematic of the RSFF in Fig. \ref{Fig: RSFFDiag}.
For the traditional RSFF, the outputs from each NOR gate is fed back into the other.
Therefore, we replace $I_1$ and $I_3$ with $R$ and $S$ and $I_2$ and $I_4$
with $Q$ and $Q'$ in (\ref{Eq: ThreshFcns}) and (\ref{Eq: NOR}),
\begin{equation}\label{Eq: ThreshFcns_RSFF}
\begin{split}
f(R,Q',x_1) &:= |R - Q'| + RQ'y_f(x_1),\\
g(S,Q,x_2) &:= |S - Q| - 1 + SQy_g(x_2);
\end{split}
\end{equation}
\begin{equation}\label{Eq: RSFF}
\begin{split}
Q_{n+1} &= (1-R)(1-Q'_n),\\
Q'_{n+1} &= (1-S)(1-Q_n).
\end{split}
\end{equation}
It should be noted that (\ref{Eq: ThreshFcns_RSFF})
must be used in conjunction with (\ref{Eq: Thresh}).

\subsection{Comparison with experiments}\label{Sec: Comparison}

From the models of the circuit we can simulate the NOR gate and RSFF operations.
The codes for the simulations are quite simple with the majority of it dedicated to
iterating (\ref{Eq: Thresh}) for both operations.  For both the NOR gate and RSFF,
we assume there is a ``circuit frequency'', which we define as the amount of time
it takes a signal to traverse the entire circuit, and is on the order of $100$ microseconds.
We also set the following parameters: $\mu_f = 3.2$, $\nu_f = 0.5694$,
$\mu_g = 2$, and $\nu_g = 1/3$.

\begin{figure}[htbp]
\centering
\includegraphics[width = 0.45\textwidth]{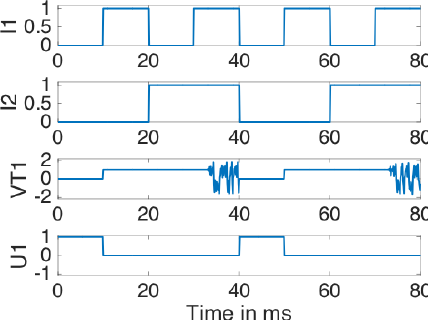}\quad
\includegraphics[width = 0.45\textwidth]{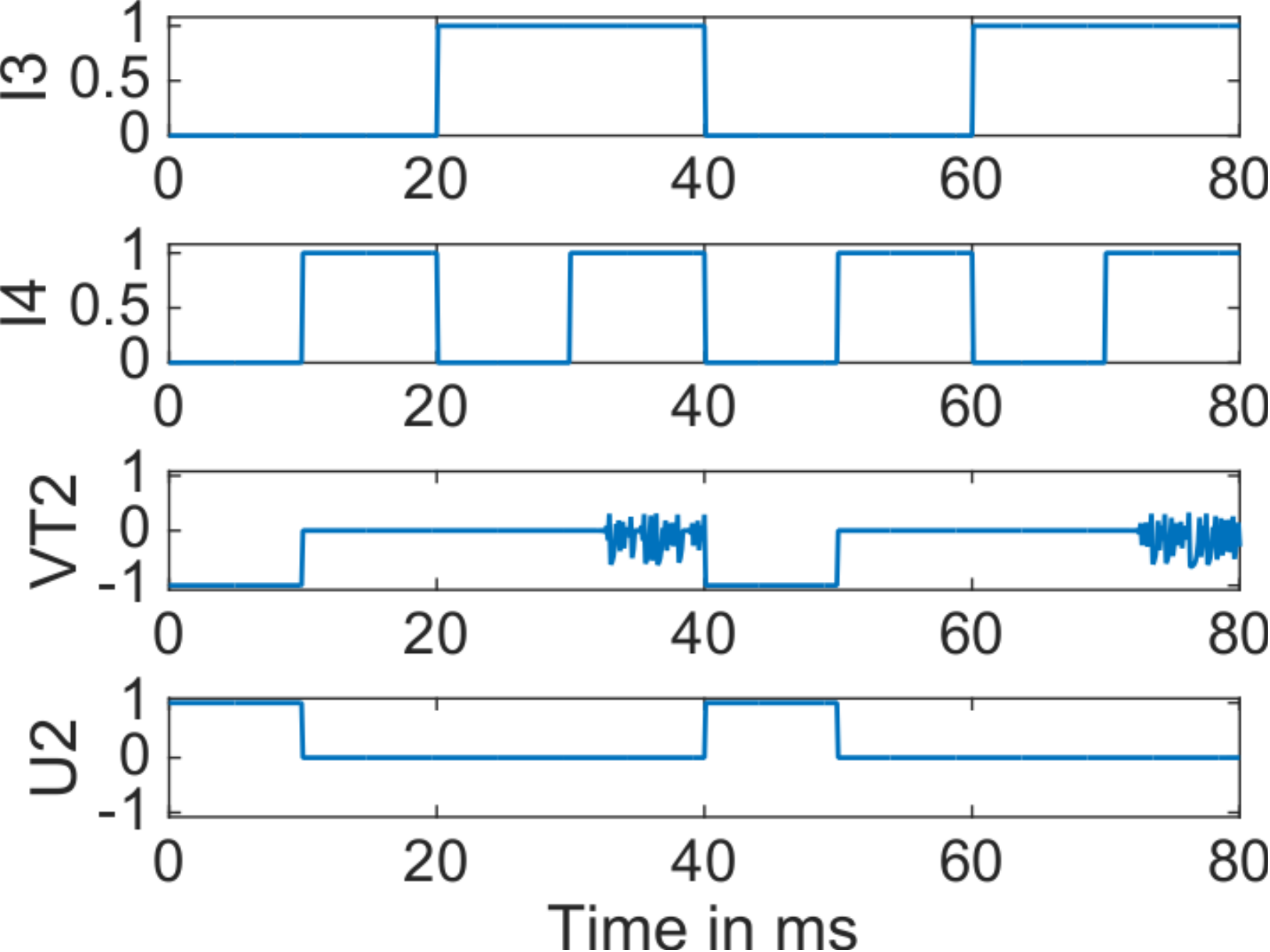}
\caption{Simulations of the two input voltages, threshold voltage, and output voltages respectively
for NOR gate operations.}\label{Fig: NOR}
\end{figure}

Furthermore, for the NOR gate, while the model
is not stochastic (no added noise), we do make small deterministic perturbations in
the inputs to approximate the effects of noise and demonstrate
sensitivity to initial conditions.  For $I_2$ and $I_4$, from the $20$ millisecond
mark to the $40$ millisecond mark we add $10^{-9}$, and from the $60$ millisecond
mark to the $80$ millisecond mark we subtract $10^{-9}$.  This equates to less
than a nano-volt difference (recall the voltage used in the experiments is $1.84$
volts).  The simulations are plotted in Fig. \ref{Fig: NOR}.

Notice, we have surprisingly close agreement with Fig. \ref{Fig: NOR_ExpResults}.
The model even replicates the lag observed in the second threshold
(in Fig. \ref{Fig: NOR_ExpResults} the second threshold voltage seems to remain
close to zero for a few milliseconds).
The only dynamics that were missed are the effects on the outputs at the
clock edges.  This shall be rectified in the sequel.

\begin{figure}[htbp]
\centering
\begin{subfigure}[t]{0.3\textwidth}
\centering
\includegraphics[width = \textwidth]{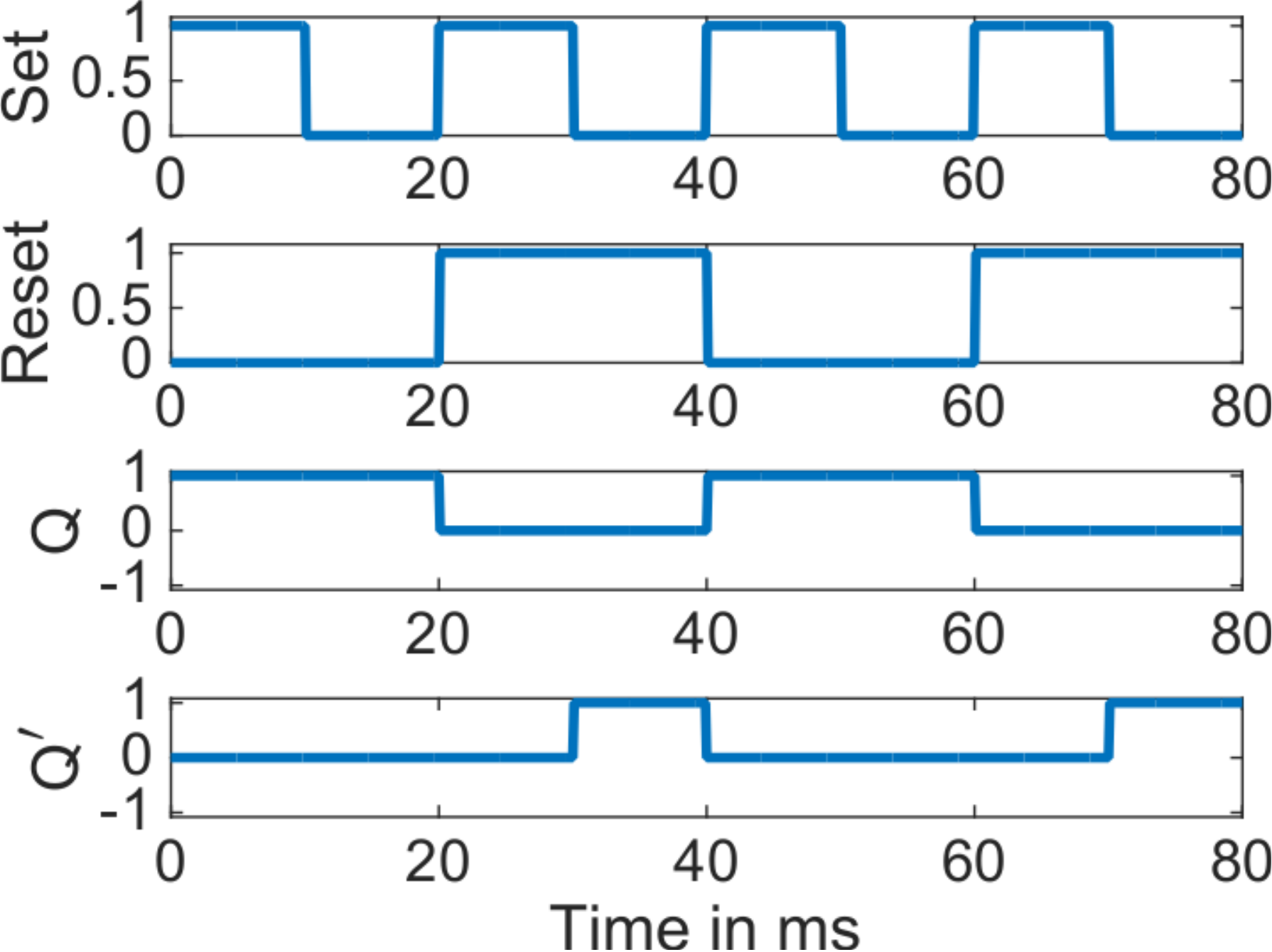}
\caption{Plots of two input voltages and simulation of two output voltages respectively for RSFF operations.}
\label{Fig: RSFF}
\end{subfigure}
\qquad
\begin{subfigure}[t]{0.61\textwidth}
\centering
\includegraphics[width = 0.49\textwidth]{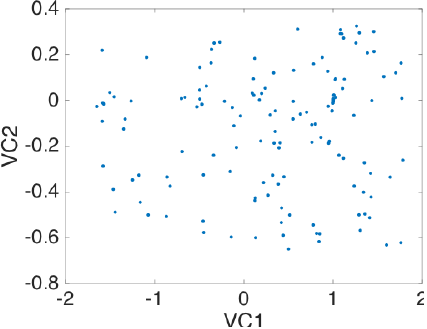}
\includegraphics[width = 0.49\textwidth]{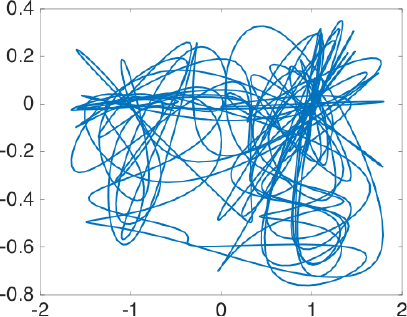}
\caption{Simulation of the capacitor voltages.  On the left we plot the individual points
from the model.  On the right we interpolate between these plots using $2000$ points
between each two iterates.}\label{Fig: Capacitors}
\end{subfigure}
\caption{Simulation of RSFF operation.}
\end{figure}

For the RSFF, we employ the same circuit frequency and perturbation on the
initial conditions.  This is plotted in Fig. \ref{Fig: RSFF}.
Moreover, in the circuit design, the outputs are achieved by taking the difference
between the voltage across the capacitors and their respective threshold voltages.
However, we approach this from the other direction where we have a model for
the outputs and threshold voltages and take the sum to predict the voltage across
the capacitors.  Since our system is discrete and the phase plane of Chua's circuit
is continuous, we interpolate between the respective points.  While this is not perfectly
accurate, it illustrates a more complete picture of the dynamics than the purely
discrete case.  This is plotted in Fig. \ref{Fig: Capacitors}.

Again, as with the NOR gate, we are able to reproduce the behavior for the RSFF
except for the clock edge effects.  We also plot the iterate plane for the capacitor
voltages.  This can be thought of as iterates of a Poincar\'{e} map of the double scroll
attractor.  After interpolating between the iterates, we observe a double scroll like
projection in the plane.  However, it is not exactly a double scroll projection, nor can
it be due to the artificial interpolation.

\section{Stochastic model for set/reset flip-flop}\label{Sec: Stochastic}

\subsection{Derivation}

Now let us rectify the discrepancies between Figs. \ref{Fig: NOR}, \ref{Fig: RSFF} and
Figs. \ref{Fig: NOR_ExpResults}, \ref{Fig: RSFF_ExpResults}.  Physically, the oscillations observed at
certain clock edges are caused by a non-binary difference in the capacitor voltage
and threshold voltage (recall that the outputs are calculated by subtracting the
threshold voltage from the capacitor voltage).  However, we take a different approach,
developing a model for the threshold voltages and output voltages rather than
determining the output voltages via the other two as explained in Sec. \ref{Sec: Deterministic}.
In order to accomplish this, we must still rely on the physical intuition of the
thresholding mechanism.

Whenever the inputs are such that they cause a threshold to change its state
drastically (not including the gradual transition to chaos), the TCU attempts
to synchronize the capacitor voltage with the threshold voltage for the proper
output.  This causes a competition between the capacitor voltage (under the
influence of the previous input) and the TCU (stimulated by the current input),
which the TCU finally wins.  During this process, due to the chaotic nature of
both the capacitor voltages and threshold voltages, each path to synchronization
is different.  Since we do not have the explicit model for the capacitor voltages,
we will treat this competition as a stochastic process.

During the transitions that lead to the edge effect, we assume there is a
probability at each time step that the output will either accept the new
inputs or be induced by the weighted average of inputs from previous time
steps, where the weights are also determined randomly (or rather,
pseudo-randomly).  Here we define time step as the reciprocal of the circuit
frequency.

Consider the sequences $\lbrace m \rbrace_1^M$, $\lbrace n\rbrace_0^N$,
and $\lbrace t\rbrace_0^T$, such that $T = N+1$
and $M = T$.  Let $N$ be the number of time steps in the past that affect
future outcomes and $T$ be the number of time steps after the edge that the
output is affected (usually about $1$ millisecond).  Further, let
$\left\lbrace w_j^{(t)},\ldots,w_j^{(t-n+1)},w_j^{(t-n)},w_j^{(t-n-1)},\ldots,w_j^{(t-N)} \right\rbrace
\subseteq \lbrace m\rbrace_1^M$ be the weights applied, with the same probability,
out of $\lbrace m\rbrace_1^M$,
to the respective inputs, $I_j^{(t-n)}$ for $j = \lbrace 1,2,3,4\rbrace$.  Let $p(t) = \lbrace 0,1\rbrace$ be
some random variable for each time $t$, with not necessarily identical distributions.
In addition, let
\begin{equation}\label{Eq: Perceived}
A_j^{(t)} = \begin{cases}
I_j^{(t)} &\text{ if } p(t) = 0,\\
\left( \sum_{n=0}^N w_j^{(t-n)}I_j^{(t-n)} \right) \bigg/ \sum_{n=0}^N w_j^{(t-n)}
&\text{ if } p(t) = 1;
\end{cases}
\end{equation}
be the perceived input.  Then we substitute $A_j^{(t)}$ into (\ref{Eq: NOR}) for the respective inputs,
\begin{equation}
\begin{split}
U_1(t) &= \left(1-A_1^{(t)}\right)\left(1-A_2^{(t)}\right),\\
U_2(t) &= \left(1-A_3^{(t)}\right)\left(1-A_4^{(t)}\right);
\end{split}
\end{equation}
to model the effects at the transition points.  It should be noted that besides
the transition points, which last about a millisecond, the outputs behave as
described by the continuous extensions.  Finally, In order to simulate noise
produced by the wave generator, we add/subtract a small random term,
$\epsilon \in \R$, to the inputs (i.e. $I_j \rightarrow I_j + \epsilon$).

For the RSFF, we use the same technique, except this time we plug (\ref{Eq: Perceived})
in only for the $R$ and $S$.  Hence, for the transitions, (\ref{Eq: RSFF}) becomes,
\begin{equation}
\begin{split}
Q_{n+1}(t) &= \left(1-A_1^{(t)}\right)\left(1-Q'_n\right),\\
Q'_{n+1}(t) &= \left(1-A_3^{(t)}\right)\left(1-Q_n\right).
\end{split}
\end{equation}

\subsection{Comparison with experiments}

\begin{figure}[htbp]
\centering
\includegraphics[width = 0.45\textwidth]{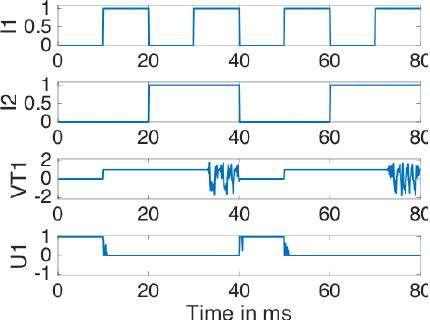}\quad
\includegraphics[width = 0.45\textwidth]{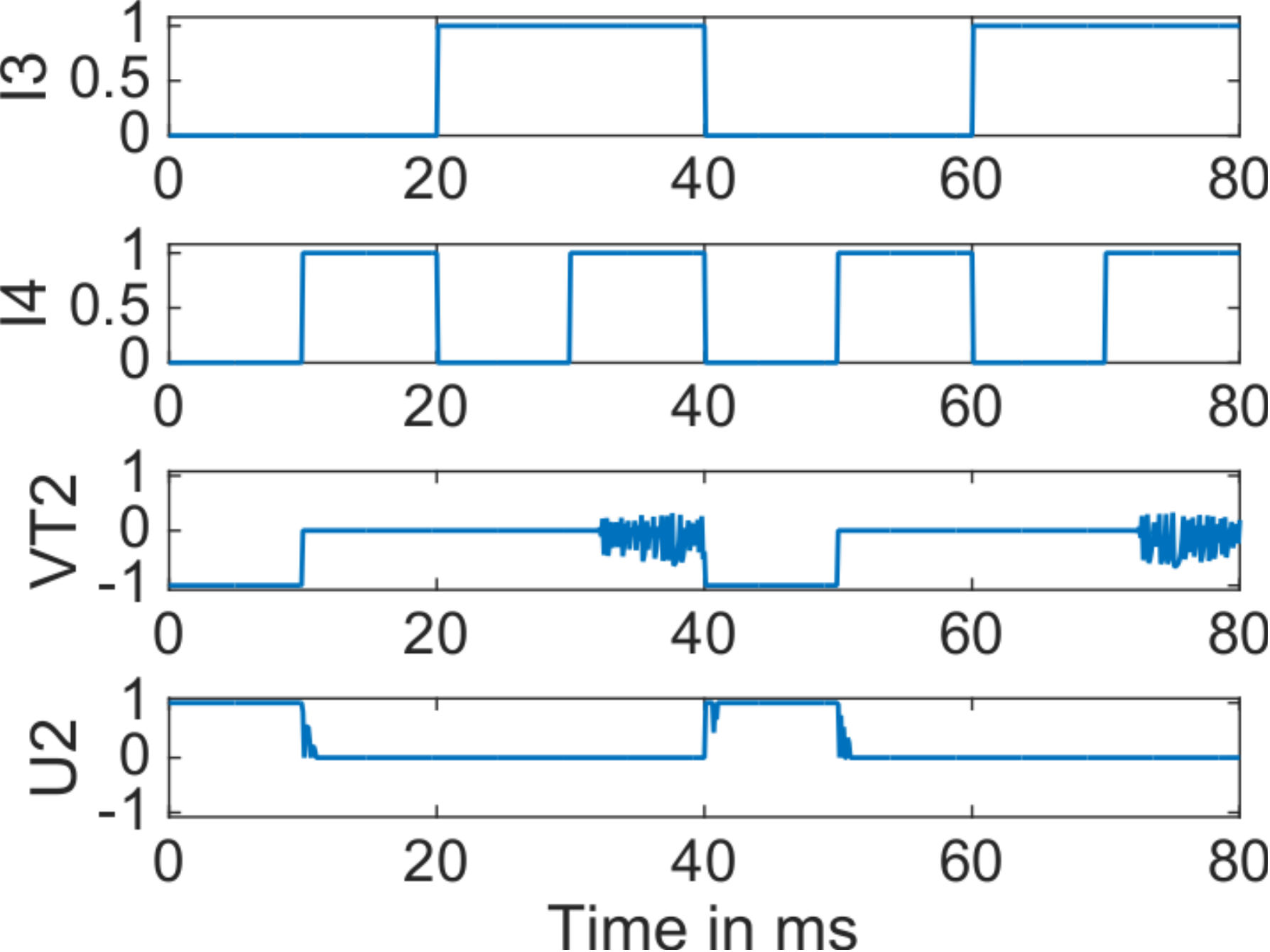}
\caption{Stochastic simulations of the two input voltages, threshold voltage, and output voltages respectively
for NOR gate operations.}\label{Fig: Stochastic_NOR}
\end{figure}

To simulate the models we use the same values for parameters in $y_f$ and $y_g$,
and for the circuit frequency, as Sec. \ref{Sec: Comparison}.  For the NOR gate and RSFF we
set $M = N + 1 = T = 11$ (i.e. approximately $1$ millisecond) and $\epsilon = \pm O\left(10^{-8}\right)$.
The choice of $\epsilon$ describes a physical signal which has less than $1$ nano-volt
of noise, on average.  The simulations for the NOR gate are plotted in Fig. \ref{Fig: Stochastic_NOR}.
We observe that these are precisely the type of damped oscillations seen in Fig. \ref{Fig: NOR_ExpResults}.

\begin{figure}[htbp]
\centering
\begin{subfigure}[t]{0.3\textwidth}
\centering
\includegraphics[width = \textwidth]{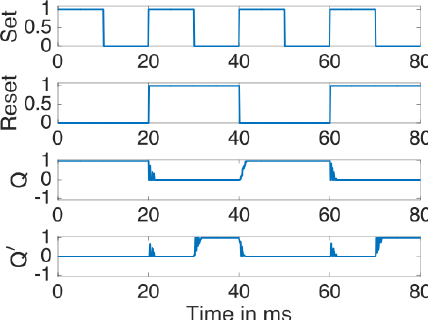}
\caption{Plots of two input voltages and simulation of two output voltages respectively for RSFF operations.}
\label{Fig: Stochastic_RSFF}
\end{subfigure}
\qquad
\begin{subfigure}[t]{0.61\textwidth}
\centering
\includegraphics[width = 0.49\textwidth]{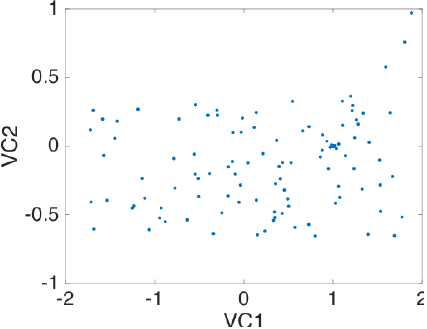}
\includegraphics[width = 0.49\textwidth]{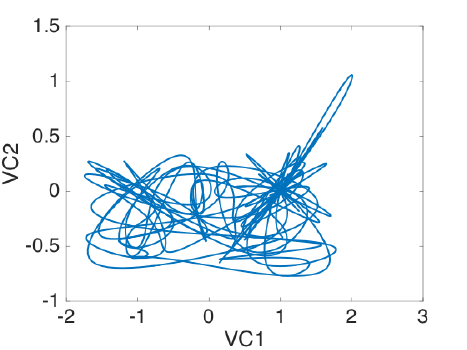}
\caption{Stochastic simulation of the capacitor voltages.  On the left we plot the individual points
from the model.  On the right we interpolate between these plots using $2000$ points
between each two iterates.}\label{Fig: Stochastic_Capacitors}
\end{subfigure}
\caption{Simulation of RSFF operation.}
\end{figure}

The RSFF and capacitor voltage simulations are plotted in Fig. \ref{Fig: Stochastic_RSFF}
and \ref{Fig: Stochastic_Capacitors} respectively.  Observe that the oscillations match the
type in Fig. \ref{Fig: RSFF_Output}.  Furthermore, Fig. \ref{Fig: Stochastic_Capacitors} is now
qualitatively more similar to Fig. \ref{Fig: Exp_Double_Scroll}.

\pagebreak

\section{Conclusions}\label{Sec: Conclusion}

In the context of chaotic logical circuits there has been many \emph{SPICE} simulations, some experiments,
and only a few models.  Simpler chaotic logical circuits are modeled as ordinary differential equations
\cite{ONK1998, Kacprzak88}.  More complex ones are generally studied using \emph{SPICE} simulations
\cite{MSD2003A, MSD2003B, Cafagna-Grassi06} and physical realizations \cite{MSD2003A, MSD2003B}.
We studied the RSFF/dual NOR gate through experiments, dynamical modeling, simulations, and analysis
of the models.

By modifying the circuit in \cite{Cafagna-Grassi06} we designed an RSFF/dual NOR gate using
modern components.  We then simulated the circuit using \emph{MultiSIM} to confirm the new design
produces the proper outputs.  Next we put together a physical realization of the circuit and
conducted experiments to show agreement with \emph{MultiSIM}.  By observing the behavior of the
circuit and using properties of TCUs as seen in \cite{MSD2003A, MSD2003B} we are able to
model the dynamics of the circuit as difference equations.  The chaotic behavior of the models
are verified by standard dynamical systems analysis for one-dimensional maps.  Finally, we
simulate our models to show agreement with both experiments and \emph{MultiSIM}.  It was expected
that the original deterministic models would not be sufficient to replicate the ``race'' behavior
observed in the outputs.  Therefore, we inserted probabilistic elements to show this edge-trigger
phenomenon, thereby deriving a stochastic model.

While we are able to capture much of the behavior, there is a need for more sophisticated models
to replicate the more complex ``race'' behavior.  Furthermore,
as this is a rich problem, and we have a physical realization, many new phenomena may arise.
It shall be useful to study various physical bifurcations as we make changes in the circuit and make
connections with topological bifurcations.  We predict this will lead to local bifurcations such as
transcritical, pitchfork, and Neimark--Sacker as observed in other models \cite{BRS2009, RahmanBlackmore16},
and novel global bifurcations previously unobserved in the literature.  We shall also endeavor to
build other more complex circuits to study analogs of other logic families and exploit the chaotic
and logical properties for the purposes of encryption and secure communication.

\section{Acknowledgement}

The authors would like to give their sincere thanks to the NJIT URI phase-2 student seed grant
for funding this project.  D. Blackmore and A. Rahman appreciates the support of
DMS at NJIT, and I. Jordan appreciates the support of ECE at NJIT.  The authors also thank
Parth Sojitra for looking over the circuit when a bug was present.


\bibliographystyle{unsrt}
\bibliography{RSFF}
\nocite{GuckenheimerHolmes83, Devaney86, Strogatz94, Robinson95,
Wiggins2003, Marc}

\begin{appendices}

\section{\emph{MultiSIM} and Physical Realization}

\begin{table}[htbp]
\centering
\setlength{\tabcolsep}{12pt} 
\renewcommand{\arraystretch}{1}
\begin{tabular}{l||c|c|c|}
Type & Quantity & Code & Comments\\
\hline
$1 k\Omega$ Resistor & 9 & & \\
$100 k\Omega$ Resistor & 14 & & \\
$1.6 k\Omega$ Resistor & 2 & & \\
$22 k\Omega$ Resistor & 2 & & \\
$220 k\Omega$ Resistor & 2 & & \\
$2.2 k\Omega$ Resistor & 1 & & \\
$3.3 k\Omega$ Resistor & 1 & & \\
\hline
$100 nF$ Capacitor & 1 & & \\
$10 nF$ Capacitor & 1 & & \\
\hline
$18 mH$ Inductor & 1 & & \\
\hline
Op-Amp & 6 & AD713JN & Used only in \emph{MultiSIM}\\
Op-Amp & 1 & LM759CP & Used only in \emph{MultiSIM}\\
\hline
Op-Amp & 7 & NTE858m & Used only in physical realization\\
\hline
\end{tabular}
\caption{List of components.}
\label{Tab: Parts}
\end{table}

\begin{figure}[htbp]
\centering
\includegraphics[width = \textwidth]{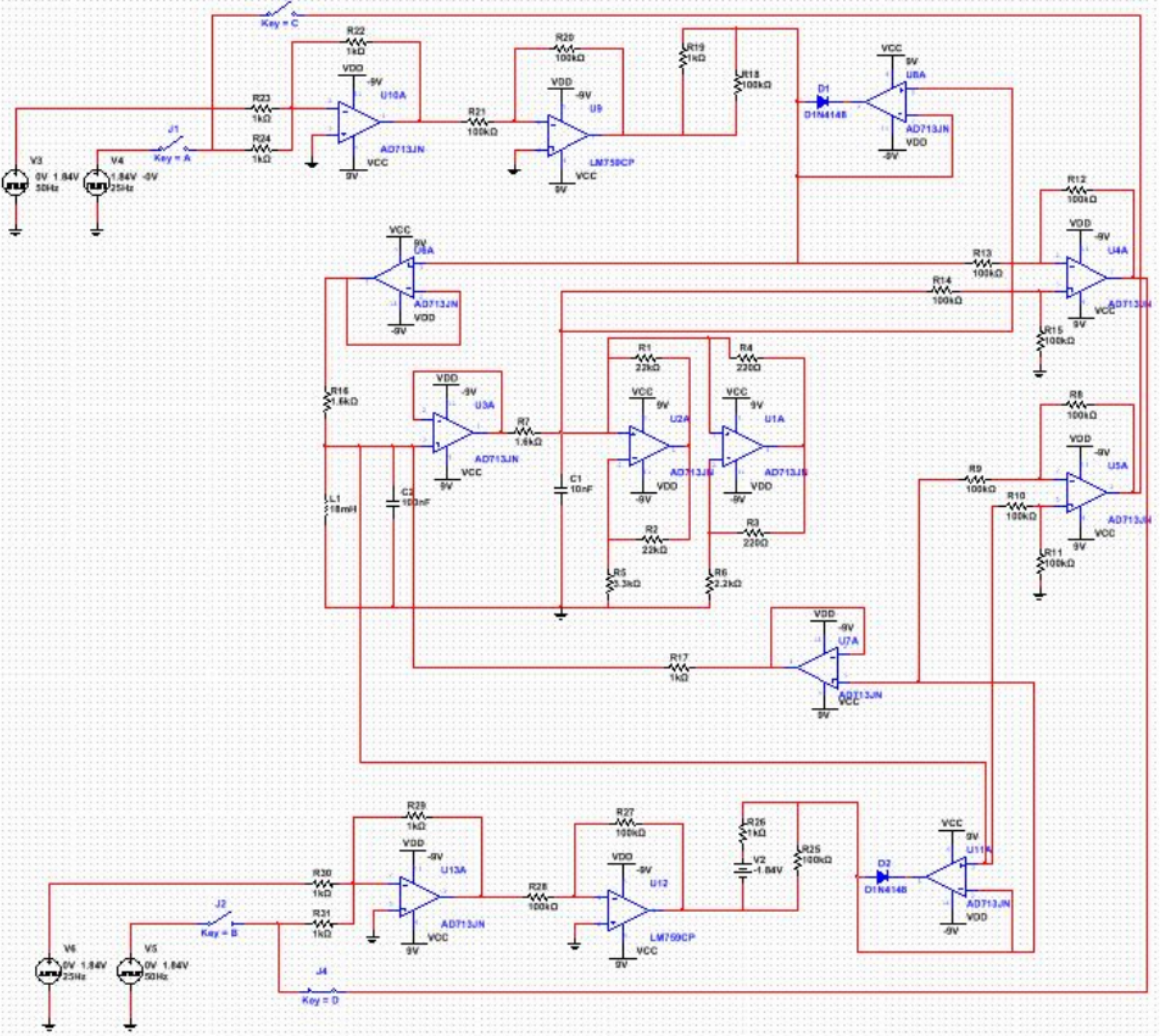}
\caption{Full schematic of the RSFF/dual NOR design.}\label{Fig: RSFFschem}
\end{figure}

\begin{figure}[htbp]
\centering
\includegraphics[width = 0.45\textwidth]{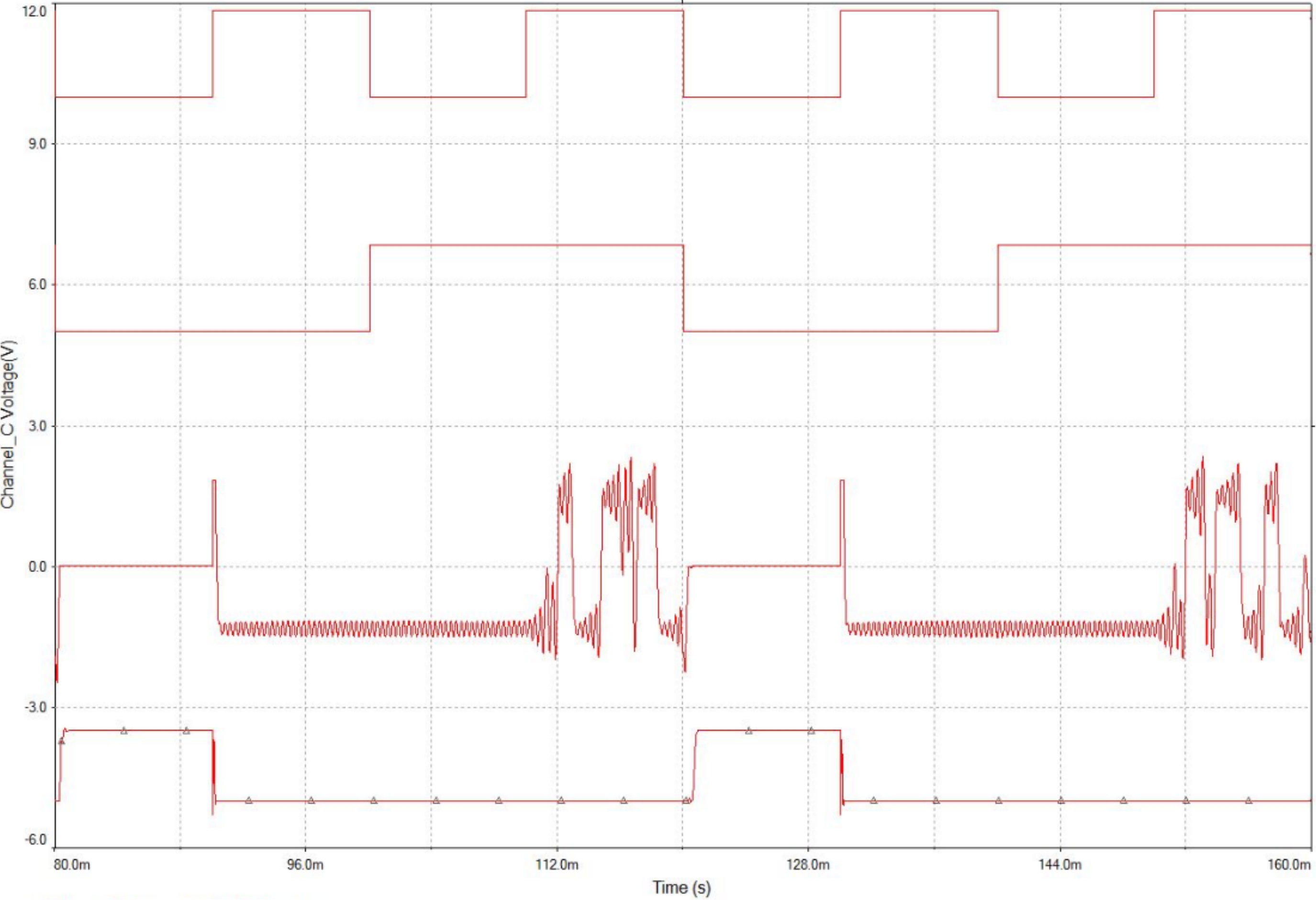}\quad
\includegraphics[width = 0.45\textwidth]{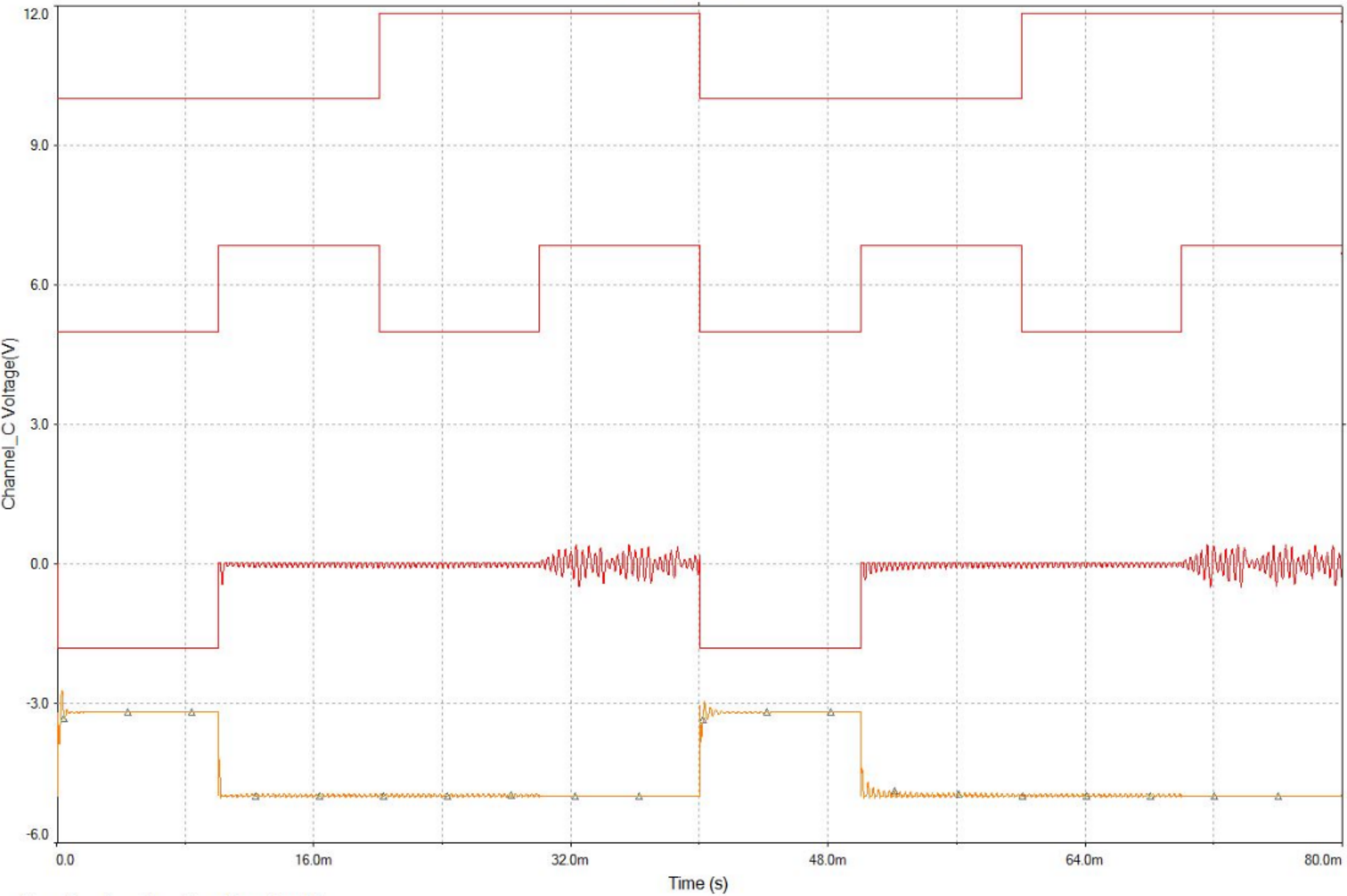}
\caption{\emph{MultiSIM} plots of the two input voltages, threshold voltage, and output voltages respectively
for the two separate NOR gates.}\label{Fig: Ian_NOR}
\end{figure}

\begin{figure}[htbp]
\includegraphics[width = \textwidth]{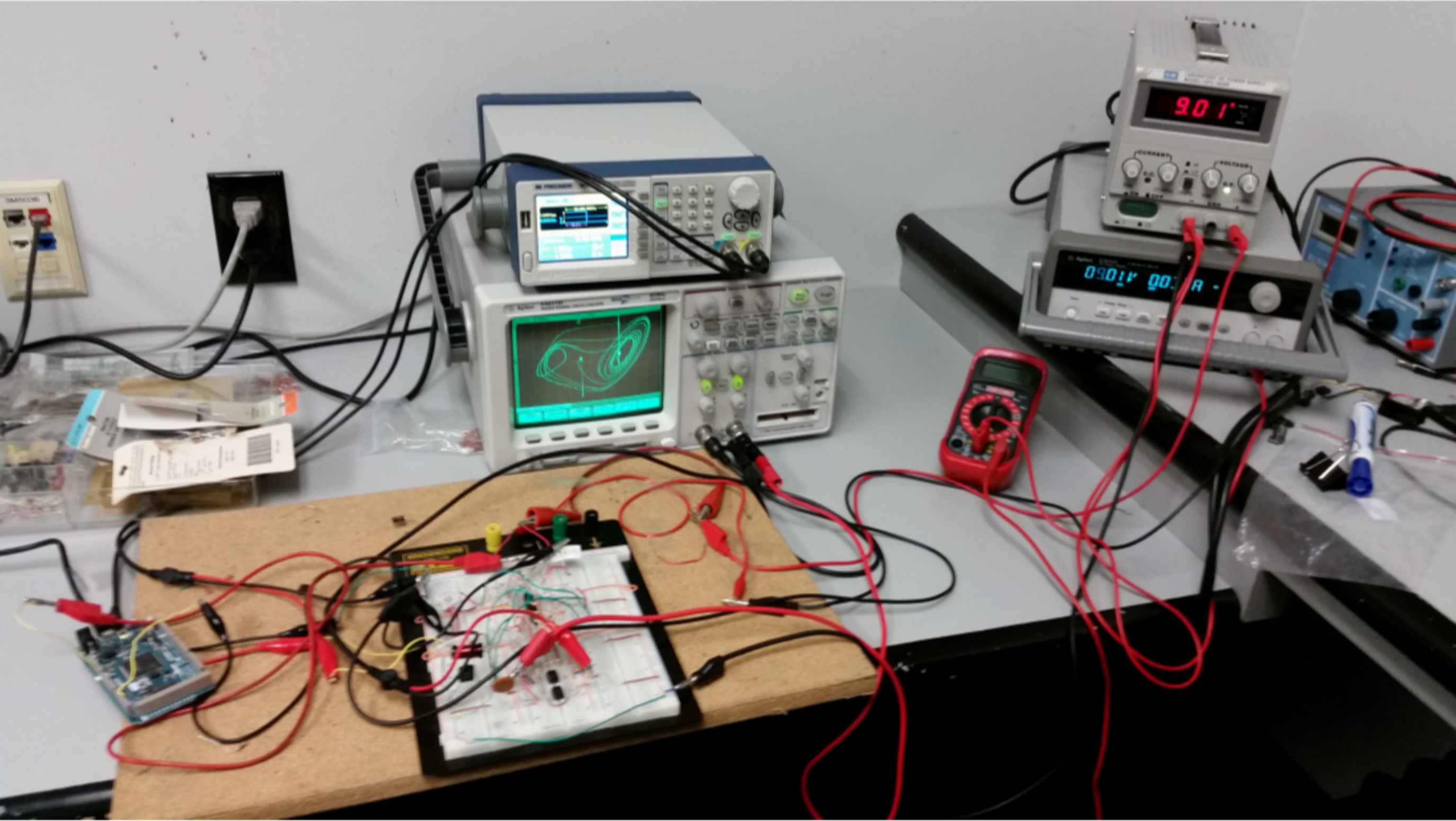}
\caption{Experimental setup}\label{Fig: ExpSetup}
\end{figure}

\end{appendices}

\end{document}